\documentclass[a4paper,11pt]{article}
\pdfoutput=1
\usepackage[utf8]{inputenc}
\usepackage{amssymb}
\usepackage{amsmath}
\usepackage{amsthm}
\usepackage{amsfonts}
\usepackage[pdftex]{graphicx}
\usepackage{geometry}
\usepackage{braket}
\usepackage{enumerate}
\usepackage{comment}
\usepackage{lscape}
\usepackage{graphicx}
\usepackage{amssymb}
\usepackage{amsmath}
\usepackage{bbm}
\usepackage{cancel}
\usepackage{epstopdf}
\usepackage{mathtools}
\usepackage{braket}
\usepackage[export]{adjustbox}
\usepackage[usenames,dvipsnames]{xcolor}
\usepackage{tikz}
\usetikzlibrary{cd}
\usepackage{relsize}
\usepackage{titlesec}
\usepackage{setspace} 
\usepackage{xpatch}
\usepackage{lipsum} 
\usepackage{xfrac}
\usepackage[final]{hyperref}
\hypersetup{
    colorlinks=true,
    linkcolor=blue,
    urlcolor=blue,
    linktoc=all
}

\usepackage{setspace}
\usepackage[in]{fullpage}
\usepackage{hyperref}
\usepackage{array}
\newcolumntype{P}[1]{>{\centering\arraybackslash}p{#1}}
\newcolumntype{M}[1]{>{\centering\arraybackslash}m{#1}}
\usepackage{cancel}
\usepackage{wrapfig}
\usepackage[font=small,labelfont=bf]{caption}
\usepackage{pifont}

\usepackage[dvipsnames]{xcolor}
\usepackage{mathtools}

\usepackage{color}

\usepackage{tikz}
\usetikzlibrary{arrows,shapes}
\usetikzlibrary{trees}
\usetikzlibrary{matrix,arrows} 
\usetikzlibrary{positioning}
\usetikzlibrary{calc,through}
\usetikzlibrary{decorations.pathreplacing}
\usepackage{pgffor}	

\numberwithin{equation}{section}

\usepackage[english]{babel}

\usepackage{caption}

\usepackage[nottoc,notlot,notlof]{tocbibind}
\usepackage[nosort]{cite}   
\usepackage{color}

\usepackage{parskip}
\usepackage[T1]{fontenc}

\usepackage{lmodern}
\usepackage{yfonts}
\usepackage{bbm}
\usepackage{physics}
\usepackage{slashed}
\usepackage{makeidx}
\usepackage[vcentermath]{youngtab}
\usepackage{amsthm}
\usepackage[scr=boondoxo]{mathalfa}

\hypersetup{
	colorlinks=true,
	linktoc=page,
	citecolor=blue,
	linkcolor=blue,
	urlcolor=blue} 

\makeatletter \def\@captype{figure} \makeatother 
\usetikzlibrary{calc}

\makeatletter

\newcommand{\nn}{\nonumber}

\begin{document}

\begin{flushright}
	 QMUL-PH-25-30\\
\end{flushright}

\vspace{20pt} 

\begin{center}

	{\Large \bf  {Higher spin fields and the field strength multicopy} }  \\
	\vspace{0.3 cm}

	\vspace{20pt}

	{\mbox {\sf  \!\!\!\! Graham~R.~Brown* and Bill~Spence**
	}}
	\vspace{0.5cm}
		{\small \em\\
        
          *{Higgs Centre for Theoretical Physics\\
          School of Physics and Astronomy\\
          University of Edinburgh\\
          EH9 3FD,  United Kingdom\\} 
\vspace{0.5cm}
            **Centre for Theoretical Physics\\
			Department of Physics and Astronomy\\
			Queen Mary University of London\\
			Mile End Road, London E1 4NS, United Kingdom
		}

	\vspace{40pt}  

	{\bf Abstract}
		
  \end{center}
  
 We discuss the generalisation of the Weyl double copy to higher spin "multi-copies", showing how the natural linearised higher spin field strengths can be related to sums of powers of the Maxwell tensor. The tracelessness of the field strength involves the appropriate Fronsdal equations of motion for the higher spin field. 
  We work with spacetimes admitting Kerr-Schild coordinates and give a number of examples in different dimensions. We note that the multi-copy is particularly transparent in four dimensions if one uses spinor descriptions of the fields, relating this to the Penrose transform. The higher-dimensional spinor multicopy is also explored and reveals some  interesting new features arising from the little group based identification of higher spin field strengths and Maxwell tensor types.
  We then turn to the vector superspace formalism describing higher spin and `continuous' spin representations given by Schuster and Toro, based on symmetric tensor fields. Here the Kerr-Schild higher spin fields we have used earlier naturally package into a simple expression involving an arbitrary function, when the continuous spin scale $\rho$ is set to zero. Further, we discuss the case of an anti-de Sitter background, where there is also a vector space formalism given by Segal and we clarify this approach using a different definition of the covariant derivative. We give a general solution of Kerr-Schild type and  finally we describe some of the obstacles to a continuous spin formulation.

\vspace{0.3cm}
 
\noindent
 
\vfill
\hrulefill
\newline
\vspace{-1cm}
\!\!{\tt\footnotesize\{graham.brown@ed.ac.uk, w.j.spence@qmul.ac.uk\}}

\setcounter{page}{0}
\thispagestyle{empty}
\newpage


\setcounter{tocdepth}{4}
\hrule height 0.75pt
\tableofcontents
\vspace{0.8cm}
\hrule height 0.75pt
\vspace{1cm}
\setcounter{tocdepth}{2}

\section{Introduction}
\label{IntroductionSection}

The double copy has proven a fertile source of new ideas and results over recent years. It was first applied to scattering amplitudes, initially relating amplitudes in Yang-Mills theory and Gravity \cite{Bern:2008qj,Bern:2010ue}, where it continues to influence progress in increasingly diverse areas (see, for example, the SAGEX series \cite{Travaglini:2022uwo} for comprehensive reviews). Subsequently it has been explored in the context of field theory solutions, where similar progress has ensued 
(see, for example,  the book \cite{White:2024pve} and references therein). 
One particularly interesting approach to the double copy is the Weyl double copy, first explored in \cite{Luna:2018dpt}\footnote{With some related earlier work in the general relativity literature \cite{WalkerPenrose, Hughston, Dietz} }. Here the Weyl tensor in four dimensions is related to an expression quadratic in the Maxwell tensor. Subsequent work has expanded on this in a number of different directions \cite{Godazgar:2020zbv, Sabharwal:2019ngs, Alawadhi:2019urr, Alawadhi:2020jrv, Elor:2020nqe, Chacon:2021wbr, Chacon:2021lox}.

In this paper we would like to explore the generalisation of the Weyl double copy to higher spin fields, that we will  call the field strength multicopy. This  relates the higher spin $s>2$ field strengths to products of the Maxwell tensor. We will develop this in various dimensions and for both tensor and spinor formulations. We will also 
explore higher spin fields in the continuous spin formulation of \cite{Schuster:2014hca}, with a related Kerr-Schild construction, applying this to anti-de Sitter space.

We begin in section \ref{sec:higherspinKS} with the definition of the higher spin Kerr-Schild fields and the Fronsdal equations of motion, and give the recursion relation satisfied by these. 
We then turn in section \ref{sec: FSmulticopy} to define the field strength and gauge invariance for the spin $s$ fields, relating its trace and divergence to the Fronsdal equations. This allows us to formulate the field strength multi-copy, which relates the spin $s$ field strength to sums of powers of $s$ copies of the Maxwell tensor which share the same symmetry and trace properties. We give a number of examples - the Schwarschild and Eguchi-Hanson metrics in four dimensions, the Tangherlini metric in five and six dimensions, and the Myers-Perry singly rotating solution in five dimensions.

In section \ref{sec: SpinorMulticopy} we describe the spinor description of the multicopy. This avoids many of the complications of the tensor analysis and  provides new insights. In four dimensions, the spin $s$ multicopy is easily derived using the Penrose transform. This also provides simple formul\ae\ for  "mixed" multicopies
whereby the spin $s$ spinor field strength is expressed in terms of field strengths of lower spin. This includes a simple recursion relation linking the spin $s$ field strength to the product of the spin $s-1$ field strength and the Maxwell spinor. After a brief review of the spinor version of the Petrov-type classification of spacetimes, we describe how the multicopy leads to the definition of higher spin  quantities which are linked to those in the classification of Maxwell spinors. This then relates the types of higher spin field strengths which can be generated from the differing Maxwell types via the multicopy. We then explore higher-dimensional versions of the spinor multicopy, where the little group is non-trivial and plays an important role in the analysis.
We discuss the five-dimensional case, and after summarising briefly the recently discovered Petrov-type classification, we describe how the spinor multicopy relates the higher spin little group-valued quantities to the Maxwell spinors. This again then relates the types of higher spin field strengths which can be generated from the differing Maxwell fields via the multicopy, albeit in a more involved way due to the presence of the little group.

In section \ref{sec: EHBackground} we discuss the vector superspace formulation of higher spin and `continuous' spin particles (CSPs).
Here the Kerr-Schild higher spin field that we have used earlier can be written in a unified compact form involving an arbitrary function.  When the continuous spin parameter is zero we show that this unified higher spin Kerr-Schild field satisfies the vector superspace equations of motion.
Finally in this section we study the AdS-Kerr case, presenting gauge invariant higher spin field strengths defined using the relevant Kerr-Schild vector and scalar, and noting examples of the tensor multicopy.  The vector superspace formulation of the equations of motion and gauge invariance are then given, using a new covariant derivative. A generalised Kerr-Schild type solution is also given, and the simple extension of this formulation to a continuous spin particle which applied in flat space  is shown not to yield a CSP theory in this case.

In section \ref{sec: conclusion} we present conclusions and note some further avenues of exploration that our work suggests. Finally, appendix \ref{app: cotangent} contains additional details about the covariant derivative used in section \ref{sec: EHBackground}.

%
\section{Higher spin Kerr-Schild fields}\label{sec:higherspinKS}
The study of metrics  admitting a Kerr-Schild formulation \cite{KerrSchild1, KerrSchild2} (see also the review \cite{MacCallum})   has had many applications in general relativity. 
These metrics have more recently played a key role in the discovery of the double copy for
classical field theory and gravitational solutions \cite{Monteiro:2014cda, Luna:2015paa}\footnote{Some earlier work on higher spin fields and the multicopy for Kerr-Schild spacetimes is \cite{Didenko:2008va, Didenko:2009td}.}.
Here we will discuss higher spin versions of the standard Kerr-Schild fields and their equations of motion, before turning to generalisations of the Weyl double copy that can be defined based on these.

Given a static vacuum Kerr-Schild metric solution 
\begin{equation}
\label{eq:KSmetric}
  g_{\mu\nu}=\eta_{\mu\nu} +
  k_\mu k_\nu \phi\, ,
\end{equation}
with a Kerr-Schild vector $k_\mu$ satisfying
\begin{equation}\label{eq:KSvec}
  k^\nu\partial_\nu k^\mu = 0 = k^\mu k_\mu\, ,
\end{equation}
 then one can define the zeroth, single and double copy fields
\begin{equation}
\begin{split}\label{eq:SCDC}
  \phi & \, , \\
  \phi_\mu &= k_\mu \phi\, , \\
  \phi_{\mu\nu}&= k_\mu k_\nu \phi\, .
\end{split}
\end{equation}
which satisfy the spin zero, one and two equations
\begin{equation}
\begin{split}\label{eq:spin12}
  \Box\phi &=0 \, , \\
  \Box\phi_\mu -\partial^\nu\partial_\mu \phi_\nu & = 0 \, , \\
   \Box \phi_{\mu\nu}-2\partial^\rho\partial_{(\mu} \phi_{\nu)\rho} +  \partial_\mu\partial_\nu {\phi^\rho}_\rho  & = 0  \, .
 \end{split}
\end{equation}

 It is natural to then consider  the higher-spin Kerr-Schild field  
\begin{equation}\label{eq:KSphi}
\phi^{KS}_{\mu_1\dots\mu_s}:= k_{\mu_1}\dots k_{\mu_s}\phi
\end{equation}
 and show that it satisfies a higher spin linearised field equation i.e. the Fronsdal equation. This has been done in \cite{Didenko:2022qxq} (with earlier work in four dimensions in \cite{Didenko:2008va}). Note that $\phi^{KS}$ is traceless as the vector $k_\mu$ is null. To summarise their result, we define an expression whose vanishing is the Fronsdal equation for a symmetric double traceless spin $s$ field $\phi_{\mu_1\dots\mu_s}$
\begin{equation}\label{eq:Fronsdal}
 F^{(s)}_{\mu_1\dots\mu_s}(\phi^{(s)}) := \Box\phi_{\mu_1\dots\mu_s} -s\,
\partial^\lambda\partial_{(\mu_1}\phi_{\mu_2\dots\mu_s)\lambda} + \frac{1}{2}s(s-1)\partial_{(\mu_1}\partial_{\mu_2}{\phi_{\mu_3\dots\mu_s)\lambda}}^\lambda \, ,
\end{equation}
((anti-)symmetrisations are defined with unit weight e.g. $A_{(\mu\nu)}=\frac{1}{2}(A_{\mu\nu}+A_{\nu\mu})$). The Fronsdal equation is invariant under the gauge transformation
\begin{equation}\label{eq:FronsdalGinv}
\delta\phi^{(s)}_{\mu_1\dots\mu_s} = \partial_{(\mu_1}\xi_{\mu_2\dots\mu_s)}
\, 
\end{equation}
where $\xi$ is required to be traceless.

A helpful discussion of the Fronsdal equation can be found in the recent review of higher spin theories in \cite{Ponomarev:2022vjb}. The subject has a long and varied history; two earlier reviews are \cite{Sorokin:2004ie, Bouatta:2004kk}. 

We will also define $\tilde F^{(s)}_{\mu_1\dots\mu_s}$ by
\begin{equation}\label{eq:FronsdalKS}
 \tilde F^{(s)}_{\mu_1\dots\mu_s} := F^{(s)}_{\mu_1\dots\mu_s}(\phi^{KS})\end{equation}
ie the expression on the right-hand side of eqn. 
\eqref{eq:Fronsdal} with the spin $s$ Kerr-Schild
field  $\Phi^{KS}$ replacing $\phi^{(s)}$.

In ref. \cite{Didenko:2022qxq},
it is  shown that if the Fronsdal equations $\tilde F^{(s)}=0$ are satisfied for $s=0,1,2$ (i.e., eqns. \eqref{eq:spin12}) then they are satisfied for all values of $s$. This argument relies on the following recursion relation, implicit in their derivation,
\begin{equation}
\begin{split}
\tilde F^{(s)}_{\mu_1\dots\mu_s} &= k_{(\mu_1} \tilde F^{(s-1)}_{\mu_2\dots\mu_s)} 
+ (s-1) k_{(\mu_1}\dots k_{\mu_{s-2}} \tilde F^{(2)}_{\mu_{s-1}\mu_s)} 
 - (2s-3) k_{(\mu_1} \dots k_{\mu_{s-1}} \tilde F^{(1)}_{\mu_s)} \\ &
\qquad+ (s-2) k_{\mu_1} \dots k_{\mu_s} \tilde F^{(0)} \, .
\end{split}
\end{equation}
Applying this repeatedly then leads to the desired result
\begin{equation}
\tilde F^{(s)}_{\mu_1\dots\mu_s} =  \frac{1}{2}s(s-1)k_{(\mu_1} \dots k_{\mu_{s-2}} \tilde F^{(2)}_{\mu_{s-1}\mu_{s)}} 
 - s(s-2)  k_{(\mu_1} \dots k_{\mu_{s-1}} \tilde F^{(1)}_{\mu_{s)}} 
+  \frac{1}{2}(s-1)(s-2) k_{\mu_1} \dots k_{\mu_s} \tilde F^{(0)} \, .
\end{equation}
%


\section{The field strength multicopy}\label{sec: FSmulticopy}

Now consider field strengths associated to the above higher spin fields. The linearised field strength associated to a symmetric higher spin field $\phi_{\mu_1}\dots _{\mu_s}$ is the $2s$-component tensor
\begin{equation}\label{eq:FieldStrength}
H^{(s)}_{\mu_1\nu_1\mu_2\nu_2\dots\mu_s\nu_s}(\phi^{(s)}) = \partial_{\mu_s}\partial_{\mu_{s-1}}\dots\partial_{\mu_1} \phi^{(s)}_{\nu_1\nu_2\dots\nu_s}\Big\vert_{[\mu_i\nu_i]}\, ,
\end{equation}
where the notation $\vert_{[\mu_i\nu_i]}$ means to antisymmetrise the expression in each pair of indices $(\mu_i,\nu_i)$, for $i=1,...,s$.
The field strength is invariant under the gauge transformations
\begin{equation}\label{eq:GaugeTrans}
\delta\phi^{(s)}_{\mu_1\dots\mu_s} = \partial_{(\mu_1}\Lambda_{\mu_2\dots\mu_{s)}}\, .
\end{equation}
for a symmetric spin $s-1$ field $\Lambda$.\footnote{A discussion of field strengths, gauge symmetries, (non-local) actions and the equivalents of the Einstein tensor for fields in arbitrary representations of the Lorentz group can be found in \cite{deMedeiros:2002qpr, deMedeiros:2003osq}}

It is straightforward to show that the trace of the field strength is related to the Fronsdal equation by 
\begin{equation}\label{eq:TraceH} 
\eta^{\mu_1\nu_s} H^{(s)}_{\mu_1\nu_1\mu_2\nu_2\dots\mu_s\nu_s}
= -\frac{1}{4}\partial_{\mu_2}\dots\partial_{\mu_{s-1}}      F^{(s)}_{\nu_2\dots\nu_{s-1}\nu_1\mu_s}\Big\vert_{[\mu_i\nu_i]} \, ,
\end{equation}
where the antisymmetrisation on the right-hand side is over each pair $(\mu_i,\nu_i)$ for $i=2,\dots,s-1$. 
The divergence of the field strength is also related to the Fronsdal equation, by the equation
\begin{equation}\label{eq:TraceH2} 
\partial^{\mu_1} H^{(s)}_{\mu_1\nu_1\mu_2\nu_2\dots\mu_s\nu_s}
= \frac{1}{2}\partial_{\mu_2}\dots\partial_{\mu_s}     F^{(s)}_{\nu_2\dots\nu_s\nu_1}\Big\vert_{[\mu_i\nu_i]} \, ,
\end{equation}
 where here again the antisymmetrisation on the right-hand side is over each pair $(\mu_i,\nu_i)$ for $i=2,\dots,s$.
If we insert the Kerr-Schild solution for the spin $s$ field into these equations we thus deduce that the field strength for this field is traceless and divergence-free.

Now consider the linearised Weyl double copy. This concerns examples where the linearised Weyl tensor $C^{(0)}_{\mu\nu\rho\sigma}$ is proportional to a sum of terms quadratic in the Maxwell field strength tensor. Explicitly 
\begin{equation}\label{eq:WeylDC} 
C^{(0)}_{\mu\nu\rho\sigma} = \alpha \Big(F_{\mu\nu}F_{\rho\sigma} - F_{\rho\mu}F_{\nu\sigma} -\frac{6}{d-2} \eta_{\mu\rho}{F_\nu}^\lambda F_{\sigma\lambda}+ \frac{3}{(d-1)(d-2)}\eta_{\mu\rho}\eta_{\nu\sigma}F^{\lambda\pi}F_{\lambda\pi}\Big)\Big\vert_s \, ,
\end{equation}
where $\alpha$ is a proportionality factor, $d$ is the number of spacetime dimensions and the symbol $\vert_s$ here means to antisymmetrise in the pairs of indices $(\mu,\nu)$ and $(\rho,\sigma)$. Note that both sides of this equation are traceless using the flat metric. Here the Maxwell and Weyl tensor are just special cases of the higher spin field strength tensors defined in \eqref{eq:FieldStrength}
\begin{equation}
    F\coloneqq 2H^{(1)}_{\mu\nu}(\phi^{(1)})\,, \quad  C^{(0)}_{\mu_1\nu_1\mu_2\nu_2}\coloneqq 2 H^{(2)}_{\mu_1\nu_1\mu_2\nu_2}(\phi^{(2)})\,.
\end{equation}

One can then consider "multi-copy" versions of the linearised Weyl double copy, where the spin $s$ field strength can be expressed in terms sums of products of $s$ Maxwell tensors which share the same symmetry and trace properties. For example, a triple copy where the field strength for a spin three field, given by \eqref{eq:FieldStrength} above for $s=3$, is proportional to an expression cubic in the Maxwell tensor which has the same symmetries - 
\begin{equation}\label{eq:triplecopy} 
H^{(3)}_{\mu\nu\rho\sigma\lambda\tau}(\phi^{(3)}) \propto C[F]_{\mu\nu\rho\sigma\lambda\tau}:=F_{\mu\nu}F_{\rho\sigma}F_{\lambda\tau}  +\dots  \, ,
\end{equation}
where the dots indicate terms which ensure the symmetries and tracelessness of the field strength on the left-hand side of the equation are reproduced by the combination of Maxwell tensors. 

To see this explicitly, define
\begin{equation}\label{eq:Fcubedguy} 
X_{\mu\nu\rho\sigma\lambda\tau}= ( F_{\mu\nu}F_{\rho\sigma}F_{
\lambda\tau
} + 3F_{\mu\nu}F_{\rho \lambda}F_{\sigma\tau}+2F_{\mu \rho}F_{\nu \lambda}F_{\sigma \tau})\vert_{symm}\ ,
\end{equation}
and the trace terms (taken with the flat metric)
\begin{equation}\label{eq:F3traces} 
X'_{\mu\nu\rho\sigma\lambda\tau}= \eta_{\mu\nu}{X^\alpha}_{\rho \alpha\sigma \lambda \tau}, \quad X''_{\mu\nu \rho \sigma \lambda \tau}=  \eta_{\mu\nu}\eta_{\rho\sigma}{X^{\alpha\beta}}_{\alpha \lambda\beta \tau}\ ,
\end{equation}
where the notation $\vert_{symm}$ means to antisymmetrize the expression in the indices $(\mu\nu),(\rho\sigma),(\lambda\tau)$ and symmetrize under the interchange of any two of those pairs. The expression in \eqref{eq:Fcubedguy} vanishes upon antisymmetrisation in any three adjacent indices.
Then the triple copy expression is given explicitly by
\begin{equation}
\label{eq:triplecopyexplicit}
\begin{split}
C[F]_{\mu\nu\rho\sigma\lambda\tau} = & X_{\mu\nu\rho\sigma\lambda\tau} - \frac{12}{d}(X'_{\mu\rho\nu \sigma\lambda\tau} + X'_{\mu\rho\lambda\tau \nu \sigma}+X'_{\lambda\tau \mu \rho\nu \sigma})\vert_{symm}\\ & \!\!\!\!\!\!\!\!\!\!\!\!\!\!\!
- \frac{12}{d(d+1)}\Big(X''_{\nu \rho\mu \sigma\lambda\tau}+X''_{\nu \rho\lambda\tau \mu \sigma}+X''_{\lambda\tau \nu \rho\mu \sigma} - 2(X''_{\lambda \mu \tau \sigma\nu \rho}+X''_{\lambda \mu \nu \rho\tau \sigma}+X''_{\nu \rho\lambda \mu \tau \sigma})\Big)\vert_{symm}
\, . 
\end{split}
\end{equation}


\subsection{Schwarzschild metric}
To give a simple example, consider the Schwarzschild metric, written in the Kerr-Schild form
\begin{equation}\label{eq:Schwarzmetric}
ds^2 = (\eta_{\mu\nu} + h_{\mu\nu}  )dx^\mu dx^\nu= (\eta_{\mu\nu} + \frac{\lambda}{r}k_\mu k_\nu  )dx^\mu dx^\nu\, ,
\end{equation}
with $k_\mu=(1,x/r,y/r,z/r)$ and $\lambda$ a constant.

First consider the linearised double copy. The field strength is given by
\begin{equation}\label{eq:linDC} 
C^{(0)}_{\mu\nu\rho\sigma} = 2\partial_\mu\partial_\rho h_{\nu\sigma} \vert_s\, ,
\end{equation}
and the Maxwell single copy gauge field is
\begin{equation}\label{eq:MaxSC} 
A_\mu = \frac{\lambda}{r}k_\mu\, .
\end{equation}
The linearised double copy is then
\begin{equation}\label{eq:linearisedDC} 
C^{(0)}_{\mu\nu\rho\sigma} = \frac{2r}{\lambda} \Big(F_{\mu\nu}F_{\rho\sigma} - F_{\rho\mu}F_{\nu\sigma} -3 \eta_{\mu\rho}{F_\nu}^\lambda F_{\sigma\lambda}+ \frac{1}{2}\eta_{\mu\rho}\eta_{\nu\sigma}F^{\lambda\pi}F_{\lambda\pi}\Big)\Big\vert_s \, ,
\end{equation}
where indices are raised with the flat metric in the expression above.

For the triple copy, define the spin three field
\begin{equation}\label{eq:spin3} 
\phi_{\mu\nu\rho} = \frac{\lambda}{r}k_\mu k_\nu k_\rho\, , 
\end{equation}
and its field strength
\begin{equation}\label{eq:spin3FS} 
H^{(3)}_{\mu_1\nu_1\mu_2\nu_2\mu_3\nu_3} = \partial_{\mu_1} \partial_{\mu_2} \partial_{\mu_3}  \phi_{\nu_1\nu_2\nu_3}\vert_{[\mu_i,\nu_i]}\, , 
\end{equation}
where the notation indicates that the expression on the right-hand side is to be antisymmetrised in the pairs $(\mu_i,\nu_i)$ for $i=1,2,3$. This field strength is traceless with the flat metric.

The triple copy is then found to be
\begin{equation}\label{eq:lindTC} 
H^{(3)}_{\mu_1\nu_1\mu_2\nu_2\mu_3\nu_3} = \frac{5r^2}{2\lambda^2} C[F]_{\mu_1\nu_1\mu_2\nu_2\mu_3\nu_3} \, . 
\end{equation}
%


\subsection{Eguchi-Hanson}\label{sec:EH}
Another example is the Euclidean Eguchi-Hanson metric. With coordinates $(u,v,X,Y)$ the Kerr-Schild vector is $k_\mu = \frac{1}{uv-XY}(v,0,0,-X)$ and the flat metric and metric perturbation are given by
\begin{equation}\label{eq:EHh} 
 \eta_{\mu\nu}dx^\mu dx^\nu= 2(du dv - dX dY)\,,\quad h_{\mu\nu} = \frac{m}{uv-XY}k_\mu k_\nu\,, 
\end{equation}
with $m$ a constant, and the total metric given by
\begin{equation}
    ds^2_{\text{EH}}=(\eta_{\mu\nu}+h_{\mu\nu})dx^\mu dx^\nu\,.
\end{equation}
The Maxwell gauge field is
\begin{equation}\label{eq:EHMax} 
A_\mu  = \frac{m}{uv-XY}k_\mu\, . 
\end{equation}

The linearised field strength is given again by
\begin{equation}\label{eq:linDEH} 
C^{(0)}_{\mu\nu\rho\sigma} = 2\partial_\mu\partial_\rho h_{\nu\sigma} \vert_s ,
\end{equation}
and the double copy is 
\begin{equation}\label{eq:linearisedDC_EH} 
C^{(0)}_{\mu\nu\rho\sigma} = \frac{uv-XY}{m} C[F]_{\mu\nu\rho\sigma}  \, .
\end{equation}
For the triple copy, define the spin three field
\begin{equation}\label{eq:spin3EH} 
\phi_{\mu\nu\rho} = \frac{m}{uv-XY}k_\mu k_\nu k_\rho, .
\end{equation}
and its field strength 
\begin{equation}
H^{(3)}_{\mu_1\nu_1\mu_2\nu_2\mu_3\nu_3} = \partial_{\mu_1} \partial_{\mu_2} \partial_{\mu_3}  \phi_{\nu_1\nu_2\nu_3} \vert_{[\mu_i,\nu_i]}, 
\end{equation}
and we find the triple copy expression
\begin{equation}\label{eq:lindTC_EH} 
H^{(3)}_{\mu_1\nu_1\mu_2\nu_2\mu_3\nu_3} = \frac{5(uv-XY)^2}{8m^2} C[F]_{\mu_1\nu_1\mu_2\nu_2\mu_3\nu_3} \, .
\end{equation}
%

\subsection{D>4}

For examples in more than four dimensions, 
first consider the Tangherlini metric in five dimensions. In Kerr-Schild coordinates $(t,x,y,z,w)$  the Kerr-Schild vector is given by $k_i=(1,x/r,y/r,z/r,w/r)$, where here $i=1,...,5$ and $r^2=x^2+y^2+z^2+w^2.$
The metric is
\begin{equation}\label{eq:Tangherlini}
ds^2 = (\eta_{ij} + h_{ij}  )dx^i dx^j= (\eta_{ij} + \frac{\lambda}{r^2}k_i k_j  )dx^i dx^j\, ,
\end{equation}
with $\lambda$ a constant.

For the linearised double copy, the field strength is given by
\begin{equation}
C^{(0)}_{ijkl} = 2\partial_i\partial_k h_{jl} \vert_s\, ,
\end{equation}
and the Maxwell single copy gauge field is
\begin{equation}\label{eq:MaxSC5d} 
A_i = \frac{\lambda}{r^2}k_i\, .
\end{equation}
The linearised double copy is then 
\begin{equation}\label{eq:linearisedDC2} 
C^{(0)}_{ijkl} = \frac{r^2}{\lambda} \Big(F_{ij}F_{kl} - F_{ki}F_{jl} -2 \eta_{ik}{F_j}^m F_{lm}+ \frac{1}{4}\eta_{ik}\eta_{jl}F^{mn}F_{mn}\Big)\Big\vert_s \, ,
\end{equation}
where we antisymmetrize in $(i,j)$ and $(k,l)$, and indices are raised with the flat metric in the expression above.

For the triple copy, define the spin three field
\begin{equation}
\phi_{ijk} = \frac{\lambda}{r}k_i k_j k_k\, .
\end{equation}
and its field strength
\begin{equation}
H^{(3)}_{i_1j_1i_2j_2i_3j_3} = \partial_{i_1} \partial_{i_2} \partial_{i_3}  \phi_{j_1j_2j_3}\vert_{[i_a,j_a]}\, .\end{equation}
This is traceless with the flat metric.

The triple copy is then found to be 
\begin{equation}\label{eq:lindTC2} 
H^{(3)}_{i_1j_1i_2j_2i_3j_3}= \frac{5r^4}{8\lambda^2} C[F]_{i_1 j_1i_2 j_2i_3j_3} \, . 
\end{equation}
This can be generalised to the higher-dimensional Tangherlini metrics in a straightforward way - for example the double and triple copies in six dimensions are
\begin{equation}\label{eq:linearisedDC6d} 
C^{(0)}_{ijkl} = -\frac{20 r^3}{27\lambda}\Big(F_{ij}F_{kl} - F_{ i}F_{jk} -\frac{3}{2} \eta_{ik}{F_i}^m F_{lm}+ \frac{3}{20}\eta_{ik}\eta_{jl}F^{mn}F_{mn}\Big)\Big\vert_s \, ,
\end{equation}
and
\begin{equation}\label{eq:lindTC6} 
H^{(3)}_{i_1j_1i_2j_2i_3j_3}= \frac{70 r^6}{27\lambda^2} C[F]_{i_1 j_1i_2 j_2i_3j_3}\, , 
\end{equation}
where we remind the  reader that $C[F]$, defined in \eqref{eq:triplecopyexplicit}, should be evaluated in the appropriate space time dimension. 


For a different five-dimensional example, consider the singly-rotating Myers-Perry solution. The Weyl double copy for this case was studied in a particular coordinate system in \cite{Alawadhi:2020jrv}, but here we will use  Kerr-Schild coordinates $x^i=(t,x_1,x_2,y_1,y_2)$ so we can set up the formalism for studying multiple copies.

The metric is given by
\begin{equation}\label{eq:MPmetric} 
g_{ij} = \eta_{ij} + h k_i k_j
\end{equation}
where the Kerr-Schild vector $k_i$ is given by
\begin{equation}\label{eq:MPKSvector} 
k_i=\Big(1,\frac{R x_1 - a y_1}{R^2 + a^2},\frac{x_2}{R}, \frac{R y_1 + a x_1}{R^2 + a^2}, \frac{y_2}{R}\Big) \ .
\end{equation}
The vector $k_i$ is null with respect to both the flat and full metric. The inverse metric is simply
\begin{equation}\label{eq:MPinversemetric} 
g^{ij} = \eta^{ij} -h k^i k^j \ .
\end{equation}
We raise the index on $k_i$ with either the flat or curved metric.

The parameter $a$ is the rotation constant, and the function $h$ in \eqref{eq:MPmetric} is
\begin{equation}\label{eq:MPh} 
h = \frac{\mu}{(R^2 + a^2) \bigg(1 - \frac{a^2 (x_1^2 + y_1^2)}{(R^2 + a^2)^2}\bigg)} \ ,
\end{equation}
where $\mu$ is a constant. Finally, $R$ is a solution of the equation
\begin{equation}\label{eq:MPR} 
 \frac{x_1^2+y_1^2}{R^2+a^2}    + \frac{x_2^2+y_2^2}{R^2} = 1\ .
\end{equation}

Now consider the linearised  Weyl double copy. The Maxwell field strength is given by
\begin{equation}\label{eq:Maxwell} 
F_{ij} = \partial_i(h k_j) - \partial_j(h k_i)\ .
\end{equation}
Use this to form a tensor with the same symmetries as the Weyl tensor, which is also traceless with respect to the flat metric as 
\begin{equation}\label{eq:MaxWeyl} 
C[F]_{ijkl}=\Big(F_{ij}F_{kl} - F_{ki}F_{jl}  - 2 \eta_{ik} {F_j}^s F_{ls} + \frac{1}{4} \eta_{ik}\eta_{jl}F^{st}F_{st}\Big)\Big\vert_{s}  \ ,
\end{equation}
where the notation $\vert_s$ means to antisymmetrise the expression within the brackets in $(i,j)$ and $(k,l)$.

The linearised Weyl double copy can then be written
\begin{equation}\label{eq:linWeylDC} 
C^{(0)}_{ijkl}= \frac{1}{h} C[F]_{ijkl}  \ .
\end{equation}

Now consider the triple copy. Firstly, define the spin three field strength by 
\begin{equation}\label{eq:MPspin3fieldstrength} 
H^{(3)}_{ijklst}= \partial_s\partial_k\partial_i (h k_j k_l k_t)\Big\vert_{s}  \ ,
\end{equation}
where here $\vert_s$ means to antisymmetrise the expression within the brackets in $(i,j)$, $(k,l)$ and $(s,t)$.

The field strength $H^{(3)}_{ijklst}$ has the following properties - it is antisymmetric in the pairs of indices $(i,j)$, $(k,l)$ and $(s,t)$, and symmetric under the interchange of any two of these pairs. The tensor also vanishes if any three indices are anti-symmetrised. Finally, it can be checked that it is traceless using the flat metric. We find that the triple copy in this case is given by 
\begin{equation}\label{eq:MPtriplecopy} 
H^{(3)}_{ijklst}=  -\frac{5}{8h^2}\, C[F]_{ijklst} \, .
\end{equation}

There are obvious generalisations of the tensor triple copy, studied  in various examples above, to higher spins, defining a spin $s$ field strength using $s$ derivatives of the appropriate function $h$ times $s$ copies of the Kerr-Schild vector $k_i$, and showing that this
is reproduced by the sum of products of $s$ copies of the Maxwell tensor which shares the same symmetry and trace features. This will involve an increasing number of  terms. We will see in the following sections that this is much simplified in the spinor approach, which will also allow a study of generalised Petrov types for the Maxwell, spin two and higher spin spinor field strengths, and show how multicopies lead to relations between these.


\section{The spinor multicopy}\label{sec: SpinorMulticopy}

\subsection{Four dimensions}  \label{sec: 4dPart1}
We now show how many of the above results can be derived using a spinor based approach. Firstly consider four dimensions. Then the spinor form of the Maxwell tensor is a two index symmetric spinor $\phi_{AB}$, along with its conjugate $\phi_{A'B'}$, ($A,B,A',B'=(1,2)$) and the spinor equivalent of the Weyl tensor is a four index symmetric spinor $\phi_{ABCD}$ along with its conjugate $\phi_{A'B'C'D'}$.
One can immediately construct a symmetric four index spinor from the Maxwell spinor as an object proportional to $\phi_{(A'B'}\phi_{C'D')}$. Using twistor space results, an understanding of how and under what conditions this maps between solutions of the spin one and two equations and with what proportionality factor has been given in \cite{White:2020sfn} (see also \cite{Chacon:2021wbr}), and we will use the results of that paper, and their notation here. 

Using the Penrose transform, one can derive the scalar, and electromagnetic and gravitational spacetime field strength spinors in terms of the spacetime principal spinors $\alpha_{A'}=(1,\xi_1), \beta_{B'}=(1,\xi_2)$ as
\begin{equation}
\label{eq:ZSDcopy}
   \phi = \frac{N(x)}{\xi_1-\xi_2},\quad     \phi_{A'B'} = -\frac{N^2(x)}{(\xi_1-\xi_2)^3}\alpha_{(A'}\beta_{B')}, \quad 
    \phi_{A'B'C'D'} = \frac{N^3(x)}{(\xi_1-\xi_2)^5}\alpha_{(A'}\beta_{B'}\alpha_{C'}\beta_{D')}\, , 
\end{equation}
where $N(x)$ is the normalisation factor arising from the twistor/spinor incidence relation. An advantage of the twistor approach is that it identifies this function explicitly - a simple example is the Schwarschild metric where one can derive the result $\phi\propto 1/r$ \cite{White:2020sfn}, as can be seen in the tensor equation \eqref{eq:MaxSC} earlier. Notice also that in this construction the Weyl double copy is restricted to type D spacetimes, i.e., those with two pairs of repeated principal spinors\footnote{More general double copies which yield a variety of spacetime types have been found in \cite{Chacon:2021wbr} and the Weyl double copy with sources studied in \cite{Armstrong-Williams:2024bog}.}

Using the spin $s$ Penrose transform (see eqn. (13) of \cite{White:2020sfn}), one can derive the general result 
\begin{equation}
\label{eq:WhiteMulticopy}
   \phi_{A'_1B'_1\dots A'_s B'_s} = \frac{(-1)^sN^{s+1}(x)}{(\xi_1-\xi_2)^{2s+1}}\alpha_{(A'_1}\beta_{B'_1}\dots\alpha_{A'_s}\beta_{B'_s)}\, . 
\end{equation}
Using this and the previous equation one can  immediately write down the spin $s$ multicopy

\footnote{
The spin $s$ spinor multicopy for AdS${}_4$ black holes was first studied in \cite{Didenko:2009td}, and a more recent discussion of the flat and AdS cases is in\cite{Didenko:2021vui}.}
\begin{equation}
\label{eq:Spin3multicopy}
  \phi_{A'_1B'_1\dots A'_s B'_s} = \frac{(\xi_1-\xi_2)^{s-1}}{N^{s-1}(x)}   \phi_{(A'_1B'_1}\dots\phi_{A'_sB'_s)}\, . 
\end{equation}

One can also consider equivalent expressions which could be interpreted as a "mixed" multicopy - e.g. for spin three
$\phi_{A'B'C'D'E'F'}= \frac{(\xi_1-\xi_2)^2}{N^2(x)}\phi_{(A'B'}\phi_{C'D'}\phi_{E'F')} = \frac{\xi_1-\xi_2}{N(x)}\phi_{(A'B'C'D'}\phi_{E'F')}$. A particular example then takes the form of a recursion relation
\begin{equation}
\label{eq:Spin3multicopyRR}
  \phi_{A'_1B'_1\dots A'_s B'_s} = -\frac{N(x)}{(\xi_1-\xi_2)^{2}}   \phi_{(A'_1B'_1\dots A'_{s-1}B'_{s-1}} \phi_{A'_s B'_s)}\, . 
\end{equation}

For an explicit example, consider the Eguchi-Hanson metric of section \ref{sec:EH}. 
In \cite{Luna:2018dpt} it was shown that in this case the Weyl spinor is proportional to the product of two Maxwell spinors. To connect with the discussion just above and the higher spin copies, define
the spin-$s$ spinor field strengths by
\begin{equation}
\label{eq:EHspinorfieldstrengths}
\phi_{A'_1B'_1\dots A'_s B'{_s}} = \frac{1}{2^s} {\sigma^{\mu_1\nu_1}}_{(A'_1B'_1}\dots
{\sigma^{\mu_s\nu_s}}_{A'_sB'_{s})} H_{\mu_1\nu_1\dots\mu_s\nu_s}
\, . 
\end{equation}
with
\begin{equation}
\label{eq:EHtensorfieldstrengths}
H_{\mu_1\nu_1\dots \mu_s\nu_s} = \big(\nabla_{(\mu_1}\dots\nabla_{\mu_s)}\phi_{\nu_1\dots\nu_s}- {\rm traces}\big) \big\vert_{[\mu_i,\nu_i]} 
\, , 
\end{equation}
where $\phi_{\nu_1\dots\nu_s}=k_{\nu_1}\dots k_{\nu_s}\phi$ is the Kerr-Schild spin-$s$ field and the notation indicates antisymmetrisation in the pairs $(\mu_i,\nu_i)$ for $i=1\dots s$. (Note that the spinor index symmetrisation in \eqref{eq:EHspinorfieldstrengths} projects out the traces of $H$.)
Now consider the (unscaled) Newman-Penrose  spinors\footnote{
forming the null vectors $k,n,m,\bar m$ in the usual way from these, the only non-zero innner products are $k.n=-(uv-XY)^2$ and $m.\bar m=(uv-XY)^2$.}
\begin{equation}
\label{eq:EHspinors}
\hat\alpha_{A'}=\frac{1}{\sqrt{2}}\big(u-iY,u+iY\big),\quad \hat\beta_{A'}=\frac{1}{\sqrt{2}}\big(v+iX,-(v-iX)\big)\, . 
\end{equation}
Then $\hat\alpha.\hat\beta:=\hat\alpha^{A'}\hat\beta_{A'}=-(uv-XY)$ and one can derive the results
\begin{equation}
\label{eq:EHmulti}
\begin{split}
\phi=-\frac{m}{\hat\alpha.\hat\beta},
\quad &\phi_{A'B'}=-\frac{m}{(\hat\alpha.\hat\beta)^3}\hat\alpha_{(A'}\hat\beta_{B')},\quad 
\phi_{A'B'C'D'}=-\frac{3m}{(\hat\alpha.\hat\beta)^5}\hat\alpha_{(A'}\hat\beta_{B'}\hat\alpha_{C'}\hat\beta_{D')},\\
& 
\phi_{A'B'C'D'E'F'}=-\frac{15 m}{(\hat\alpha.\hat\beta)^7}\hat\alpha_{(A'}\hat\beta_{B'}\hat\alpha_{C'}\hat\beta_{D'}\hat\alpha_{E'}\hat\beta_{F')}, \\
&
\phi_{A'B'C'D'E'F'G'H'}=-\frac{105 m}{(\hat\alpha.\hat\beta)^9}\hat\alpha_{(A'}\hat\beta_{B'}\hat\alpha_{C'}\hat\beta_{D'}\hat\alpha_{E'}\hat\beta_{F'}\hat\alpha_{G'}\hat\beta_{H')}\, , 
\end{split}
\end{equation}
leading to the conjecture that 
\begin{equation}
\label{eq:generalone}
\phi_{A'_1B'_1\dots A'_s B'{_s}} = -\frac{m(2s-1)!! }{(\hat\alpha.\hat\beta)^{2s+1}}\alpha_{(A'_1}\beta_{B'_1}\dots\alpha_{A'_s}\beta_{B'_s)}
\, .
\end{equation}
This reproduces the relations \eqref{eq:WhiteMulticopy} for the Eguchi-Hanson case, up to purely numerical factors, noting that $\alpha^A\beta_A=-(\xi_1-\xi_2)$
\footnote{These formul\ae\ also follow for the linearised case, with partial derivatives in \eqref{eq:EHtensorfieldstrengths} and the flat space sigma matrices in \eqref{eq:EHspinorfieldstrengths}, giving the spinor versions of the formul\ae\ in section \ref{sec:EH}.}.

Both the Maxwell and Weyl spinors may be assigned to different classes reflecting how algebraically special they are, using the Petrov classification (see, for example,  the discussions in \cite{Monteiro:2018xev} and \cite{Luna:2018dpt} and references therein). This can be presented using a Newman-Penrose tetrad of null vectors $k^\mu, n^\mu, m^\mu, \tilde m^\mu$ which can be written in terms of a pair of two-spinors $\iota^A, \tilde \sigma^{\dot A}$ and their conjugates. 

The Maxwell spinor $\phi_{AB}$  can then be described by the three scalars
\begin{equation}   
\label{eq:NP4dMax}
 \phi_0  := \phi_{AB}\sigma^A\sigma^B, \quad \phi_1 := \phi_{AB}\sigma^A\iota^B,\quad
  \phi_2 := \phi_{AB}\iota^A\iota^B\, , 
\end{equation}
with 
\begin{equation}   
\label{eq:NP4dMax2}
  \phi_{AB}={\phi_0}\iota_A\iota_B + 
2{\phi_1}\sigma_{(A}\iota_{B)} + {\phi_2}\sigma_A\sigma_B\, . 
\end{equation}

There are then two distinct types of non-trivial Maxwell tensor - type I where $\phi_0=0$ (which can always be arranged in four, but not higher dimensions) but $\phi_1\not = 0$, and type II, where both are zero. This can also be described in terms of spinor alignment - generically one can write $\phi_{AB}=\alpha_{(A}\beta_{B)}$ for some 
choice of spinors  $\alpha_{A}, \beta_{B}$. If these are proportional then the Maxwell spinor is of type II, and if not it is type I. Note that the three quantities $\phi_0, \phi_1,\phi_2$ depend on the choice of tetrad, so that the classification of Maxwell types is more transparent  using spinor alignment.

There is an analogous well-known classification of the Weyl spinor $\phi_{ABCD}$ in curved spacetimes. This can be described by the five complex scalars
\begin{equation}   
\label{eq:NP4dWeyl}
\begin{split}
\psi_0 & :=\phi_{ABCD}\sigma^A\sigma^B\sigma^C\sigma^D,\quad
\psi_1  :=\phi_{ABCD}\sigma^A\sigma^B\sigma^C\iota^D,\quad\psi_2  :=\phi_{ABCD}\sigma^A\sigma^B\iota ^C\iota^D,\quad \\
\psi_3 & :=\phi_{ABCD}\sigma^A\iota^B\iota ^C\iota^D,\quad \psi_4  :=\phi_{ABCD}\iota^A\iota^B\iota ^C\iota^D,\quad
\end{split}
\end{equation}
(the NP spinors $\iota^A, \sigma^A$ are the curved spacetime versions here)
with
\begin{equation}   
\label{eq:NP4dWeyl2}
  \phi_{ABCD}=\psi_0\iota_A\iota_B \iota_C\iota_D  +4\psi_1\iota_{(A}\iota_B \iota_C\sigma_{D)}+6\psi_2\iota_{(A}\iota_B \sigma_C\sigma_{D)} +4\psi_3\iota_{(A}\sigma_B \sigma_C\sigma_{D)} +\psi_4\sigma_A\sigma_B \sigma_C\sigma_D\ .
\end{equation}
There are then five different Petrov classes of Weyl spinor - type I when $\psi_0=0$, type II when $\psi_0=\psi_1=0$, type D when $\psi_0=\psi_1=\psi_3=\psi_4=0$ (here there are two distinct pairs of aligned spinors), type II when $\psi_0=\psi_1=\psi_2=0$ and type N when $\psi_0=\psi_1=\psi_2=\psi_3=0$.
As in the Maxwell case, the quantities $\psi_0,\dots,\psi_4$ depend on the choice of tetrad, and the classification of Weyl spinor types can alternatively be made by studying spinor alignment - type I when there is no alignment, type II when there is one aligned pair, type D when there are two different aligned pairs, type III when there is a triplet of aligned spinors, and type N when there is a quartet.

The standard Weyl double copy is based on two copies of the same Maxwell tensor, with $\phi_{ABCD}~\sim \phi_{(AB}\phi_{CD)}$ and this imposes restrictions on which type of Weyl spinor arises, depending on the different types of Maxwell spinor. A type II Maxwell spinor, when double copied, yields a type N Weyl spinor, for example. The double copy in this case yields examples of radiation regions of isolated gravitational systems \cite{Godazgar:2020zbv}. A type I Maxwell spinor, on the other hand, yields a type D Weyl spinor under the double copy, since this yields two pairs of aligned spinors.

The triple copy analogue has $\phi_{ABCDEF}~\sim \phi_{(AB}\phi_{CD}\phi_{EF)}$ with similar conditions arising. One has seven quantities
\begin{equation}   
\label{eq:NP4dTriple}
\begin{split}
\chi_0 & :=\phi_{ABCDEF}\sigma^A\sigma^B\sigma^C\sigma^D\sigma^E\sigma^F,\quad
\chi_1  :=\phi_{ABCDEF}\sigma^A\sigma^B\sigma^C\sigma^D\sigma^E\iota^F,\quad
\\
\chi_2  &:=\phi_{ABCDEF}\sigma^A\sigma^B\sigma^C\sigma^D\iota^E\iota^F,\quad 
\chi_3  :=\phi_{ABCDEF}\sigma^A\sigma^B\sigma^C\iota^D\iota^E\iota^F,\quad
\\
\chi_4  &:=\phi_{ABCDEF}\sigma^A\sigma^B\iota^C\iota^D\iota^E\iota^F,\quad
\chi_5:=\phi_{ABCDEF}\sigma^A\iota^B\iota^C\iota^D\iota^E\iota^F,\quad 
\\
\chi_6&:=\phi_{ABCDEF}\iota^A\iota^B\iota^C\iota^D\iota^E\iota^F
\end{split}
\end{equation}
with
\begin{equation}   
\label{eq:NP4dtriple2}
\begin{split}
  \phi_{ABCDEF}=&\chi_0\iota_{A}\iota_B \iota_C\iota_D \iota_E\iota_F  
  +6\chi_1\iota_{(A}\iota_B \iota_C\iota_D \iota_E\sigma_{F)} 
 +15\chi_2\iota_{(A}\iota_B \iota_C\iota_D \sigma_E\sigma_{F)} \\
 &+20\chi_3\iota_{(A}\iota_B \iota_C\sigma_D \sigma_E\sigma_{F)} 
  +15\chi_4\iota_{(A}\iota_B \sigma_C\sigma_D \sigma_E\sigma_{F)}   
    +6\chi_5\iota_{(A}\sigma_B \sigma_C\sigma_D \sigma_E\sigma_{F)} \\
     &+\chi_6\sigma_{(A}\sigma_B \sigma_C\sigma_D \sigma_E\sigma_{F)} 
  \ .
  \end{split}
\end{equation}

One can then define different types of spin three field strength spinors based on the vanishing of sets of the seven quantities above, or using spinor alignment - a type II Maxwell spinor will triple copy to a null type spin three field strength, with six aligned spinors, and a type I Maxwell spinor will triple copy to a field strength with two triplets of aligned spinors. 
This generalises to  multicopies in the obvious way. Spinor alignment will also come into play for multicopies here and in higher dimensions as there are only a finite number of independent spinors in each dimension.


\subsection{Higher dimensions}  \label{sec: HigherdPart1}

In more than four dimensions the spinor double copy again follows from transforming the tensor version into spinor coordinates. In five dimensions, for example, the spinor indices $A,B,..$ run from $1$ to $4$, reflecting the relationship $SO(5)\sim Sp(2)$. We then have the  Maxwell and Weyl spinors
\begin{equation}   
\label{eq:MaxWeyl5d}
  \phi_{AB}= F_{ij} {\sigma^{ij}}_{AB},\quad
  \phi_{ABCD}= C_{ijkl} {\sigma^{ij}}_{AB}{\sigma^{ij}}_{CD}\, ,
\end{equation}
using the Lorentz generators  ${\sigma^{ij}}_{AB}$. Inserting a double copy formula for $C_{ijkl}$, such as those given in examples earlier, into the second equation above then leads to the spinor double copy relationship $\phi_{ABCD}\sim \phi_{(AB}\phi_{CD)}$.

 In dimensions greater than four the classification of the Maxwell and Weyl spinors is more detailed, since the little groups are non-Abelian (c.f. \cite{Monteiro:2018xev} and references therein). In the five-dimensional case, the little group is $SU(2)$, and we will use the two-spinor indices $a,b,...=1,2$. Here the Newman-Penrose pentad can be defined using  null vectors $k^i, n^i, (i=1,...5)$, together with "polarisation" vectors $\epsilon^i_{ab}$, symmetric in $(a,b)$. This pentad can be formulated in terms of spinors $k^i_a, n^i_a$, carrying little group indices. Contracting these into the field strength spinors then yields little group-valued quantities that can be used to classify the field strengths into different types.

 For the Maxwell spinor $\phi_{AB}$ this yields the bispinors
\begin{equation}   
\label{eq:MaxBispinors}
  \phi^{(0)}_{ab}:=  \phi_{AB}{k^A}_a {k^B}_b,\quad
  \phi^{(1)}_{ab}:=  \phi_{AB}{k^A}_a {n^B}_b,\quad
  \phi^{(2)}_{ab}:=  \phi_{AB}{n^A}_a {n^B}_b \, ,
\end{equation}
with 
\begin{equation}   
\label{eq:MaxBispinorsExpansion}
  \phi_{AB} = \phi^{(0)}_{ab}{n_A}^a {n_B}^b + 2\phi^{(1)}_{ab}n_{(A}^a {k_{B)}}^b+ \phi^{(2)}_{ab}{k_A}^a {k_B}^b \, .
\end{equation}

Whilst $\phi^{(0)}_{ab}$ and $\phi^{(2)}_{ab}$ are symmetric in $(a,b)$ and in the $\bf 3$ of $SU(2)$, the bispinor  $\phi^{(1)}_{ab}$ has two irreducible components - the symmetrised piece $\phi^{(1)}_{(ab)}$ and the trace $\frac{1}{2}\epsilon_{ab}\phi_{tr}^{(1)}$.

The Petrov-like classification here is then type G when none of the little group-valued bispinors in \eqref{eq:MaxBispinors} vanish, type I when only $\phi^{(0)}_{ab}=0$, and type II when both $\phi^{(0)}_{ab}=0$ and $\phi^{(1)}_{ab}=0$. However, there is a further sub-classification of type I as $\phi^{(1)}_{(ab)}$ or $\phi_{tr}^{(1)}$ may independently vanish or not. An example is a constant electric field which has $\phi^{(1)}_{(ab)}=0$ but $\phi_{tr}^{(1)}\not=0$, whereas a constant magnetic field has $\phi^{(1)}_{(ab)}\not=0$ but $\phi_{tr}^{(1)}=0$  \cite{Monteiro:2018xev} .

For the Weyl spinor $\phi_{ABCD}$ one follows an analogous path, defining five little group-valued quantities
\begin{equation}
\begin{split}\label{eq:5dWeylLittle}
   \phi^{(0)}_{abcd}&:= \phi_{ABCD}{k^A}_a {k^B}_b{k^C}_c{k^D}_d,\quad
   \phi^{(1)}_{abcd}:= \phi_{ABCD}{k^A}_a {k^B}_b{k^C}_c{n^D}_d,  \\
\phi^{(2)}_{abcd}&:= \phi_{ABCD}{k^A}_a {k^B}_b{n^C}_c{n^D}_d,\quad
    \phi^{(3)}_{abcd}:= \phi_{ABCD}{k^A}_a {n^B}_b{n^C}_c{n^D}_d,  \\
     \phi^{(4)}_{abcd}&:= \phi_{ABCD}{n^A}_a {n^B}_b{n^C}_c{n^D}_d \, , 
\end{split}
\end{equation}
with the expansion 
\begin{equation}
\begin{split}\label{eq:5dWeylLittleExp}
   \phi_{ABCD} &:= \phi^{(0)}_{abcd}{n_A}^a {n_B}^b{n_C}^c{n_D}^d + 
   4\phi^{(1)}_{abcd}{n_{(A}}^a {n_B}^b{n_C}^c{k_{D)}}^d 
   +6\phi^{(2)}_{abcd}{n_{(A}}^a {n_B}^b{k_C}^c{k_{D)}}^d \\&+ 
   4\phi^{(3)}_{abcd}{n_{(A}}^a {k_B}^b{k_C}^c{k_{D)}}^d 
   +\phi^{(4)}_{abcd}{k_A}^a {k_B}^b{k_C}^c{k_D}^d \, .
\end{split}
\end{equation}
The little group-valued quantities in eqn. \eqref{eq:5dWeylLittle} may be further broken down into irreducible representations, and a detailed classification given of the separate types and sub-types of Weyl spinor depending on which of these vanish. One can also describe these, in complex Minkowski spacetime, in terms of spinor alignment \cite{Monteiro:2018xev}. 

For the higher spin field strengths there is a similar, albeit more complex structure of field types. For example, for spin three the field strength $\phi_{ABCDEF}$ will give rise to seven little group-valued quantities $\phi^{(q)}_{abcdef}, (q=1,...7)$ upon contracting with the $n_A^a, k_A^a$ fields, which will then generate a range of little group irreducible components and a variety of types of field depending on which of these vanish.
The double copy and multi-copy will yield relationships between the types of Maxwell spinor and higher spin field strength. The Maxwell spinor for the constant electric field mentioned above, for example, is not of the most general 
type as $\phi^{(1)}_{(ab)}=0$, and this is reflected in the result that  the only non-zero gravitational quantity is
$\phi^{(2)}_{abcd}\sim \frac{1}{2}(\epsilon_{ac}\epsilon_{bd} +\epsilon_{bc}\epsilon_{ad}){(\Tr\phi^{(1)}})^2$. Similarly, for the  Maxwell field  for the Myers-Perry solution in  \eqref{eq:Maxwell}, 
the symmetric part of $\phi^{(1)}_{ab}$ vanishes.

An analogous discussion to the above applies to the multi-copy for a spin $s$ field strength, based on the relationship $\phi_{A_1,...A_{2s}}\sim \phi_{(A_1A_2}\dots \phi_{A_{2s-1},A_{2s})}$.



\section{Vector superspace and continuous spin}\label{sec: EHBackground}

Returning now to four dimensions, the classification of massless particles includes a "continuous spin" particle (CSP) with non-zero spin Casimir $W^2=-\rho^2$, for a parameter $\rho>0$ \cite{Wigner}. In this case one has an infinite tower of helicity states which are mixed under Lorentz transformations, with the mixing depending on $\rho$. A gauge theory construction of a bosonic CSP was found in  \cite{Schuster:2014hca} (see also \cite{Rivelles:2014fsa, Rivelles:2016rwo,Bekaert:2017khg, Najafizadeh:2017tin } for reviews). This  approach uses a vector superspace, and has been generalised in various ways
\cite{Schuster:2014xja, Bekaert:2015qkt,Rivelles:2016rwo,Najafizadeh:2019mun, Najafizadeh:2021dsm }, see also \cite{Schuster:2023jgc} for references. Alternative approaches to the CSP have been presented in \cite{Metsaev:2018lth,Buchbinder:2018yoo} and the relationships between the different formulations discussed in \cite{Rivelles:2014fsa, Najafizadeh:2019mun, Alkalaev:2017hvj}. Interactions have also been explored in a number of papers \cite{Schuster:2023jgc,Schuster:2023xqa,Bellazzini:2024dco,Schuster:2024wjc,Reilly:2025lnm,Kundu:2025fsd,Metsaev:2025qkr,Kundu:2025mzm,Reilly:2025vqs}.

The vector superspace approach to the bosonic continuous spin particle uses a vector coordinate $\eta^\mu$ ($\mu=1,...,4$), and the continuous spin field
\begin{equation}\label{eq: CSfield} 
\Psi(x,\eta)=\sum_{s=0}^\infty \phi^{(s)}_{\mu_1\dots\mu_s}\eta^{\mu_1}\dots\eta^{\mu_s}\, ,
\end{equation}
with spin $s$ component spacetime fields $\phi^{(s)}_{\mu_1\dots\mu_s}$.
The CSP action, equations of motion and gauge invariances for the field $\Psi$ have been presented in \cite{Schuster:2014hca}, and further discussed in \cite{Schuster:2023xqa}. The theory contains the continuous spin parameter $\rho$, and for $\rho\rightarrow{0}$ and a partial gauge-fixing the equations of motion and symmetries reduce to the Fronsdal equations \eqref{eq:Fronsdal} for each component of $\Psi$. In the following we highlight some key points of this discussion, notably from section 2 of \cite{Schuster:2023xqa}, before applying this to the Kerr-Schild field.

The equations of motion for the case $\rho=0$ are
\begin{equation}\label{eq:STeqm} 
 \delta'(\eta^2+1)\Box\Psi -\frac{1}{2} \bigtriangleup(\delta(\eta^2+1)\bigtriangleup\Psi) = 0\, ,
\end{equation}
where $\bigtriangleup =\partial.\tilde\partial$, with $\tilde\partial_\mu:=\frac{\partial}{\partial \eta^\mu}$.
The symmetries of \eqref{eq:STeqm} are
\begin{equation}\label{eq:STeqmGT} 
 \delta_\epsilon\Psi = D\epsilon := \Big( \eta.\partial-\frac{1}{2}(\eta^2+1)\bigtriangleup\Big)\epsilon(\eta,x)\, ,
\end{equation}
 where $\epsilon$ can be expressed as an expansion of coefficient fields times powers of $\eta$. Gauge invariance follows straightforwardly using the identities
\begin{equation}\begin{split}\label{eq:identities} 
\bigtriangleup(D\epsilon)&=\Box\epsilon-\frac{1}{2}(\eta^2+1)\bigtriangleup^2\epsilon\, ,\\
\delta'(\eta^2+1)D\epsilon &= \frac{1}{2}\bigtriangleup\big(\delta(\eta^2+1)\epsilon\big)
\, .
\end{split}\end{equation}
The formulation uses certain polynomials in the $\eta$ coordinates
\begin{equation}\label{eq:Ppolys} 
P_{(n)}^{\mu_1\dots\mu_n} = 2^{n/2}\Bigg(\eta^{\mu_1}\dots\eta^{\mu_n} - \frac{n(n-1)}{4} g^{(\mu_1\mu_2}\eta^{\mu_3}\dots\eta^{\mu_n)}(\eta^2+1)\Bigg)
\, ,
\end{equation}
(to avoid potential confusion with the $\eta$ coordinates, $g_{\mu\nu}$ denotes the flat space metric in this section).
It is convenient to expand the field $\Psi$ in components as 
\begin{equation}\label{eq:PsiPhi} 
\Psi = \sum_{n\geq{0}} P_{(n)}(\eta)\phi^{(n)}(x) 
\, .
\end{equation}
This shift in the definition of the component fields has the effect of decoupling them at each level, simplifying the analysis. 
We similarly define the component fields of the gauge parameter as
\begin{equation}\label{eq:Peps} 
\epsilon = \sum_{n\geq{0}} P_{(n)}(\eta)\epsilon^{(n)}(x) 
\, ,
\end{equation}

The equations of motion for the component fields of $\Psi$ at each level are then obtained by applying the polynomial $P_{(n)}$ to the equations of motion \eqref{eq:STeqm} and integrating over the $\eta$ coordinates. The $P_{(n)}$ satisfy certain orthogonality relations which have the effect of projecting the equations of motion onto individual and independent equations for the component fields of $\Psi$. The outcome is the Fronsdal equations \eqref{eq:Fronsdal} for each component.
This requires a partial gauge-fixing which imposes a double tracelessness condition on the component fields and reduces the gauge symmetry to the standard one
\begin{equation}\label{eq:STeqmGTusual} 
 \delta_\epsilon\Phi =  \eta.\partial\epsilon(\eta,x)\, ,
\end{equation}
with the components of $\epsilon$ now traceless.
We note in passing that the $\eta$ integrals may be evaluated using the results
\begin{equation}\label{eq:etaintegrals} 
\int\!d^4\eta\, \delta(\eta^2+1) F(\eta) = \sum_{n=0}c_n \big(\tilde\partial\big)^{2n} F(\eta)\big\vert_{\eta=0}\, ,\quad
\int\! d^4\eta\, \delta'(\eta^2+1) F(\eta) = \sum_{n=0}d_n \big(\tilde\partial\big)^{2n} F(\eta)\big\vert_{\eta=0}\,
\, ,
\end{equation}
for any function $F(\eta)$, with $c_n=\frac{(-1)^n}{4^n n! (n+1)!}$ and $d_n=\frac{(-1)^n}{4^n (n!)^2}$.

Now define ${\rm E}(\Psi)$ to be the left-hand side of the equations of motion \eqref{eq:STeqm}. Then
\begin{equation}\label{eq:EqMcpts} 
{\rm E}_{(s)}^{\mu_1\dots\mu_s}(\Psi)
:=\int d^4\eta P_{(s)}^{\mu_1\dots\mu_s}{\rm E}(\Psi)
\ 
\end{equation}
are the spin $s$  Fronsdal equations of motion, and the relation \eqref{eq:KSphi} then implies that ${\rm E}_{(s)}^{\mu_1\dots\mu_s}(\Phi^{KS})$ vanishes for all $s>2$, since the equations for $s=0,1,2$ are satisfied. Thus all the components of $\rm{E}(\Phi^{KS})$ are zero and the Kerr-Schild field satisfies the equations of motion \eqref{eq:STeqm}.

We can also translate our higher spin Kerr-Schild solutions to the superspace approach 
\begin{equation}\label{eq: expField}
    \hat{\Psi}=f(k \cdot \eta)\phi\,,
\end{equation}
where $k^\mu$ and $\phi$ satisfy the properties \eqref{eq:KSvec},\eqref{eq:spin12} and $f$ is an arbitrary function. We can show that this satisfies the higher spin field equations \eqref{eq:STeqm} directly for $\rho=0$. First we rewrite the equation of motion \eqref{eq:STeqm} as follows
\begin{equation}\label{eq: ST EOM Expanded}
    \delta'(\eta^2+1)\big[\Box \Psi -(\eta\cdot \partial)\Delta\Psi\big] -\frac{1}{2}\delta(\eta^2+1)\Delta^2 \Psi=0\,.
\end{equation}
Now we find that 
\begin{equation}
    \Delta \hat{\Psi}= \partial\cdot\tilde{\partial}f(k \cdot \eta)\phi
    =f'(k \cdot \eta)\partial^\nu(k_\nu \phi)\,,
\end{equation}
where we have used the fact that $(k\cdot\partial)k_\mu{=}0$. Similarly, it is immediate that 
\begin{equation}\label{eq: direct1}
     \Delta^2 \hat{\Psi}=f''(k \cdot \eta)\partial^\mu(k_\mu \partial^\nu(k_{\nu} \phi))=f''(k \cdot \eta)\partial^\mu\partial^\nu(k_\mu k_{\nu} \phi))\, ,
\end{equation}
If we contract the flat metric $g^{\mu\nu}$ into the last relation of \eqref{eq:spin12} we find
\begin{equation}\label{eq: direct2}
    \begin{split}
        g^{\mu\nu}\big(\Box (k_\mu k_\nu \phi)&-2\partial^\rho\partial_{(\mu} (k_{\nu)} k_\rho \phi) +  \partial_\mu\partial_\nu (k^\rho k_\rho \phi)\big)=0\\
        &\implies \partial^\rho\partial^{\nu} (k_{\nu} k_\rho \phi)=0\,,
    \end{split}
\end{equation}
where we have used $k^2=0$. Comparing \eqref{eq: direct1} and  \eqref{eq: direct2} we find that $\Delta^2 \hat{\Psi}=0$. Now let us consider the first term in \eqref{eq: ST EOM Expanded} which is proportional to
\begin{equation}
    \Box \Psi -(\eta\cdot \partial)\Delta\Psi\,.
\end{equation}
This also turns out to be zero for the field \eqref{eq: CSfield}.
For the first and second terms above we find 
\begin{align}
    \Box  \hat{\Psi} &=
    f''(k \cdot \eta)\left[(\partial^\mu(k\cdot \eta))(\partial_\mu(k\cdot \eta))\phi\right]+f'(k \cdot \eta)\left[\eta^\mu \Box (k_\mu \phi)\right]\, ,\\
    (\eta\cdot \partial)\Delta\hat{\Psi}&= f''(k \cdot \eta)\left[ (\eta\cdot\partial(k\cdot\eta))(\partial^\nu(k_\nu \phi))\right]+f'(k \cdot \eta)\left[\eta\cdot\partial \partial^\nu(k_\nu \phi)\right]\,.
\end{align}
When combined the second terms in the expressions above cancel due to the spin-one equation in \eqref{eq:spin12} and we obtain
\begin{equation}\label{eq: direct3}
    \Box \hat{\Psi} -(\eta\cdot \partial)\Delta\hat{\Psi}=f(k \cdot \eta) \big[(\partial^\mu(k\cdot \eta))(\partial_\mu(k\cdot \eta))\phi-  (\eta\cdot\partial(k\cdot\eta))(\partial^\nu(k_\nu \phi))\big]\,.
\end{equation}
To simplify this, note that we can write the non-zero terms of the spin-two equation in \eqref{eq:spin12} as
\begin{align}
    \eta^\mu \eta^\nu\, \Box (k_\mu k_\nu \phi)&=2(k\cdot \eta)\eta^\mu\Box(k_\mu \phi)+2(\partial^\mu(k\cdot \eta))(\partial_\mu(k\cdot \eta))\phi\,,\\
     &\nn\\
    \eta^\mu \eta^\nu\, \, 2\partial^\rho \partial_{(\mu} (k_{\nu)}k_{\rho}\phi)&=2 (\eta\cdot\partial (\eta\cdot k))(\partial^\rho(k_{\rho}\phi))+2(\eta\cdot k)\eta\cdot\partial (\partial^\rho(k_{\rho}\phi))\,,
\end{align}
which allows us to express \eqref{eq: direct3} as 
\begin{equation}
    \begin{split}
        \Box \hat{\Psi} -(\eta\cdot \partial)\Delta\hat{\Psi}=f''(k \cdot \eta)\Bigg[\frac{1}{2}\eta^\mu \eta^\nu\,&\Big( \Box (k_\mu k_\nu \phi)-2\partial^\rho \partial_{(\mu} (k_{\nu)}k_{\rho}\phi)  \Big)-\\
        &(k\cdot \eta)\eta^\mu \Big(\Box(k_\mu \phi)-\partial_\mu (\partial^\rho(k_{\rho}\phi))\Big)\Bigg]=0\,.
    \end{split}
\end{equation}
This is zero since the terms in brackets above are exactly the spin-one and spin-two field equations from \eqref{eq:spin12}. By taking $f(k\cdot\eta)=(k\cdot\eta)^s$ we recover the result from \cite{Didenko:2008va,Didenko:2022qxq} that the Fronsdal equations are satisfied by the higher-spin copies of the Kerr-Schild fields.

Now consider the plane wave solutions discussed in \cite{Schuster:2023xqa}.  These take the form
\begin{equation}\label{eq:PW1} 
\begin{split}
\Psi_{h,l}=e^{-i l.x}\psi_{h,l}(\eta) &:= e^{-i l.x} (i \eta\cdot \epsilon_+)^h, \quad
(h \geq 0), \\ \quad
\Psi_{h,l}=e^{-i l.x}\psi_{h,l}(\eta) &:=e^{-i l.x}(-i\eta\cdot\epsilon_-)^h, \quad (h \leq 0)
\ 
\end{split}
\end{equation}
where $h$ is the helicity. We use the symbol $l$ rather than the traditional $k$ to avoid confusion with the Kerr-Schild vector $k$.
We use the null complex frame vectors $(l^\mu,q^\mu,\epsilon_+^\mu,\epsilon_-^\mu)$, where $\epsilon_-^\mu$ is the complex conjugate of $\epsilon_+^\mu$ and the non-zero inner products are $q\cdot l=1$ and $\epsilon_+\cdot\epsilon_-=2$. The fields  \eqref{eq:PW1} satisfy
\eqref{eq:STeqm} since both $\Box$ and $\bigtriangleup$ annihilate them.

One can expand a general field $\Psi(\eta,x)$ in modes as
\begin{equation}\label{eq:PWexp} 
\Psi(\eta,x) = \int \frac{d^3{\bf l}}{(2\pi)^3 2 \vert{\bf l\vert}} \sum_h \Big (a_h({\bf l})\psi_{h,l}(\eta)e^{-i l\cdot x} + {\rm c.c} \Big)\Big\vert_{l^0=\vert {\bf l\vert}}\ ,
\end{equation}
where the coefficients $a_h({\bf l})$ are given by
\begin{equation}\label{eq:FTmodes} 
a_h({\bf l}) = 2\vert{\rm l}\vert\int d^4\eta \delta'(\eta^2+1) \psi^*_{h,l}(\eta)\Psi(\eta,l)\Big\vert_{l^0=\vert{\rm \bf l}\vert}\ .
\end{equation}
As an example, for the Kerr-Schild solution  $\Phi^{KS}\coloneqq e^{k.\eta}\phi$ one finds the formul\ae\ 
\begin{equation}\begin{split}\label{eq:KSmodes} 
a_{+h}({\bf l}) &= 2\vert{\bf l}\vert  \frac{i^h}{2^h h!} \int dx e^{-i l.x}(k.\epsilon_+)^h \phi(x), \quad 
\\
a_{-h}({\bf l}) &= 2\vert{\bf l}\vert \frac{(-i)^h}{2^h h!}\int dx e^{-i l.x}(k.\epsilon_-)^h\ \phi(x)
\ ,
\end{split}\end{equation}
where $k$ here is the Kerr-Schild vector, which depends on the coordinates $x^\mu$.


\subsection{Continuous spin}
Now consider the continuous spin case where $\rho\not=0$.
Here the equations of motion and gauge symmetries are obtained by simply replacing $\bigtriangleup$ by $\bigtriangleup+\rho$ \cite{Schuster:2014hca,Schuster:2023xqa}, whence
\begin{equation}\label{eq:STeqmrho} 
 \delta'(\eta^2+1)\Box\Psi -\frac{1}{2} (\bigtriangleup+\rho)(\delta(\eta^2+1)(\bigtriangleup+\rho)\Psi) = 0\, ,
\end{equation}
and
\begin{equation}
 \delta_\epsilon\Psi = \Big( \eta.\partial-\frac{1}{2}(\eta^2+1)(\bigtriangleup+\rho)\Big)\epsilon(\eta,x)\, ,
\end{equation}
This introduces interactions between the particles of different spin.
Note that we can re-write the equations of motion \eqref{eq:STeqmrho}  as
\begin{equation}\label{eq:STeqmrho2} 
\delta'(\eta^2+1)\Big(
 \Box\Psi - \eta.\partial(\bigtriangleup+\rho) \Psi + 
  \frac{1}{2}(\eta^2+1)(\bigtriangleup+\rho)^2\Psi\Big) = 0
 \, ,
\end{equation}
where we have used the distributional identity 
\begin{equation}\label{eq:iden} 
(\eta^2+1)\delta'(\eta^2+1)=-\delta(\eta^2+1)
 \, .
\end{equation}
This also implies that $(\eta^2+1)^2\delta'(\eta^2+1)=0$.

Using the fact that the Kerr-Schild field $\Phi^{KS}$ satisfies the $\rho=0$ equations of motion, we would like to find a generalisation of this which solves \eqref{eq:STeqmrho}. 
Whilst we have not succeeded in this, we note that there are some known solutions - these are the plane waves given in \cite{Schuster:2023xqa}, which are a simple extension of the solutions \eqref{eq:PW1} given by
\begin{equation}\label{eq:PW1rho} 
\begin{split}
\tilde\Psi_{h,l}=e^{-i l.x}\tilde\psi_{h,l}(\eta) &:= e^{-i l.x} e^{-i\rho\eta\cdot q} ( i\eta\cdot \epsilon_+)^h, \quad
(h \geq 0), \\ \quad
\tilde\Psi_{h,l}=e^{-i l.x}\tilde\psi_{h,l}(\eta) &:=e^{-i l.x}  e^{-i\rho\eta\cdot q}(-i\eta\cdot\epsilon_-)^h, \quad (h \leq 0)
\ 
\end{split}
\end{equation}
where $q^\mu$ is one of the null frame vectors introduced above. These plane waves are annihilated by both $\Box$ and $\bigtriangleup +\rho$ and so solve the equations of motion \eqref{eq:STeqmrho}.

If a solution $\Psi^{(\rho)}$ of \eqref{eq:STeqmrho} has the component-field expansion 
\begin{equation}\label{eq:rhocpts} 
\Psi^{(\rho)} = \sum_{s=0}\phi^{(\rho)}_{\mu_1\dots\mu_s}\eta^{\mu_1}\dots\eta^{\mu_s}
\ ,
\end{equation}
then the Fourier modes are given by
\begin{equation}\begin{split}
\label{eq:KSmodesrho} 
a^{(\rho)}_{+h}({\bf l}) &= 2\vert{\bf l}\vert\,  i^h 
\sum_{m\geq h}\int\! dx \frac{(-1)^m}{2^m}\frac{(i\rho)^{m-h}}{(m-h)!} \,\epsilon_+^{\mu_1}\dots\epsilon_+^{\mu_h}q^{\mu_{h+1}}\dots q^{\mu_m}\phi^{(\rho)}_{\mu_1\dots\mu_m}
\\
a^{(\rho)}_{-h}({\bf l}) &= 2\vert{\bf l}\vert  (-i)^h 
\sum_{m\geq h}\int\! dx \frac{(-1)^m}{2^m}\frac{(-i\rho)^{m-h}}{(m-h)!}\, \epsilon_-^{\mu_1}\dots\epsilon_-^{\mu_h}q^{\mu_{h+1}}\dots q^{\mu_m}\phi^{(\rho)}_{\mu_1\dots\mu_m}
\, ,
\end{split}\end{equation}
where the component fields have been assumed to be traceless in the above. As  would be expected, these formul\ae\ reflect the fact that the parameter $\rho$ mediates interactions between particles of different helicities. It may be possible to take the Fourier coefficients for $\Phi^{KS}$ at $\rho=0$ in \eqref{eq:KSmodes} and multiply them by the $\rho$-dependent Fourier modes of \eqref{eq:KSmodesrho} to generate a CSP version of the Kerr-Schild solution. However, we were not able to perform the inverse Fourier transform to obtain this result.


\subsection{AdS}

We turn to consider how the discussions above may generalise to anti-deSitter (AdS) backgrounds. 
(Formulations of a continuous spin particle in AdS space using the alternative light-cone, frame-like and BRST approaches have been given in \cite{Metsaev:2018lth,Metsaev:2016lhs,Metsaev:2017ytk,Metsaev:2017myp,Metsaev:2025qkr,Khabarov:2017lth,Metsaev:2019opn,Metsaev:2021zdg,Buchbinder:2024hea,Buchbinder:2024jpt}.)

Higher spin field equations  with Kerr-Schild solutions for the AdS spacetime have been studied in \cite{Didenko:2022qxq,Didenko:2011ir}. We will use the notation of \cite{Didenko:2022qxq} in the following (including the metric with signature -+++). 
Thus consider the $d$-dimensional metric  $\bar g_{\mu\nu}$ given by
\begin{equation}\label{eq:AdSmetric}
  \bar g_{\mu\nu}dx^\mu dx^\nu = -(1-\lambda r^2)dt^2 + \frac{1}{1-\lambda r^2}dr^2 + r^2 d\Omega_{d-2}^2 
    \, ,
\end{equation}
with $d\Omega_{d-2}^2$ the metric on the unit sphere in $(d-2)$-dimensions and $\lambda$ the cosmological constant. The AdS covariant derivative $\bar\nabla_\mu$, satisfies
\begin{equation}\label{eq:AdScommutator}
   [\bar\nabla_\mu,\bar\nabla_\nu] X_\rho= 2 \lambda \bar g_{\rho[\mu}X_{\nu]} 
    \, ,
\end{equation}
for a co-vector $X_\mu$.
The higher spin equations on this spacetime are 
\begin{equation}\begin{split}\label{eq:AdS_HSeqnr}
 \bar\nabla^2\phi_{\mu_1\dots\mu_s} - & s \bar\nabla_{(\mu_1}\bar\nabla^\rho \phi_{\mu_2\dots\mu_s)\rho} + \frac{1}{2}s(s-1)\bar\nabla_{(\mu_1}\bar\nabla_{\mu_2}{\phi_{\mu_3\dots\mu_s)\rho}}^\rho \\
 &+ \lambda s(s-1)\bar g_{(\mu_1\mu_2}{\phi_{\mu_3\dots\mu_s)\rho}}^\rho + \lambda(((s-2)(d+s-3) - s))\phi_{\mu_1\dots\mu_s}=0
    \, ,
\end{split}
\end{equation}
with the double trace of $\phi$ vanishing. These  have the gauge invariance
\begin{equation}\label{eq:AdSginv}
  \delta\phi_{\mu_1\dots\mu_s}= s\bar\nabla_{(\mu_1}\Lambda_{\mu_2\dots\mu_s)}
    \, ,
\end{equation}
for a traceless gauge parameter field $\Lambda_{\mu_1\dots\mu_{s-1}}$.

One can then consider field strengths in this AdS background. Considering four dimensions for the moment, the spin one field strength is just $H_{\mu\nu}=2\bar\nabla_{[\mu}\phi_{\nu]}$, whilst for spin two the gauge invariant field strength is
\begin{equation}\label{eq:spin2AdS_FS}
   H_{\mu\nu\rho\sigma} = 4\big( \bar\nabla_\rho\bar\nabla_\mu\phi_{\nu\sigma} + \lambda \bar g_{\rho\mu}\phi_{\nu\sigma}\big) \vert_{s} ,
\end{equation}
where  $\vert_s$ here means to antisymmetrize in the pairs $(\mu,\nu)$ and $(\rho,\sigma)$. 

The Kerr-Schild higher spin solutions are given by
\begin{equation}\label{eq:KSsolns}
   \phi_{\mu_1\dots\mu_s}= k_{\mu_1}\dots k_{\mu_s}\phi 
    \, ,
\end{equation}
where the scalar field and Kerr-Schild vector are given by
\begin{equation}\label{eq:scalar_KS}
   \phi=\frac{2}{r},\quad k_\mu=\Big(1,\frac{1}{1-\lambda r^2},0,0\Big)\, .
\end{equation}
The field strength \eqref{eq:spin2AdS_FS} is then traceless and divergence-free when the solution $\phi_{\mu\nu}=k_\mu k_\nu \phi$ is inserted. 

In a similar way one can write down the gauge invariant spin-3 field strength  
\begin{equation}\label{eq:spin3AdS_FS}
   H_{\mu\nu\rho\sigma\lambda\tau} = 8\Big( \bar\nabla_{(\lambda}\bar\nabla_\rho\bar\nabla_{\mu)}\phi_{\nu\sigma\tau} + 4\lambda \bar g_{(\lambda\rho}\bar\nabla_{\mu)}\phi_{\nu\sigma\tau}\Big) \Big\vert_{s} ,
\end{equation}
where here $\vert_{s}$ means to antisymmetrise in the pairs $(\mu,\nu), (\rho,\sigma), (\lambda,\tau)$. Then
$ H_{\mu\nu\rho\sigma\lambda\tau}$ is traceless and divergence-free when the spin-3 solution $\phi_{\nu\sigma\tau}=k_\nu k_\sigma k_\tau\phi$ is inserted into its definition. These results generalise to higher spins with additional terms coming in - for example for spin-4 the invariant on-shell traceless and divergence-free field strength is given by
\begin{equation}\label{eq:spin4AdS_FS}
 H_{\mu\nu\rho\sigma\lambda\tau\alpha\beta} = 16\Big( \bar\nabla_{(\alpha}\bar\nabla_{\lambda}\bar\nabla_\rho\bar\nabla_{\mu)} + 10\lambda \bar g_{(\alpha\lambda}\bar\nabla_{\rho}\bar\nabla_{\mu)}
   +9 \lambda^2 \bar g_{(\alpha\lambda}\bar g_{\rho\mu)}\Big)\phi_{\nu\sigma\tau\beta} \Big\vert_{s} ,
\end{equation}
where here $\vert_{s}$ means to antisymmetrise in the pairs $(\mu,\nu), (\rho,\sigma), (\lambda,\tau), (\alpha,\beta)$.

When the Kerr-Schild solutions \eqref{eq:KSsolns} are inserted into these field strengths, one can derive tensor expressions for the multicopies.  For example the double copy is given by
\begin{equation}\label{eq:DC_AdS}
   H_{\mu\nu\rho\sigma}(\phi^{KS}) =  
  2 r C[H]_{\mu\nu\rho\sigma}\, ,
\end{equation}
where $C[H]_{\mu\nu\rho\sigma}$ is the expression on the right-hand side of \eqref{eq:WeylDC}, with $F_{\mu\nu}$ replaced by $H_{\mu\nu}$ and the AdS metric used throughout instead of the flat metric.
Similarly the triple copy is given by
\begin{equation}\label{eq:TC_AdS}
   H_{\mu\nu\rho\sigma\lambda\tau}(\phi^{KS}) =  
  30 r^2 C[H]_{\mu\nu\rho\sigma\lambda\tau}\, ,
\end{equation}
where $C[H]_{\mu\nu\rho\sigma\lambda\tau}$ is the expression on the right-hand side of \eqref{eq:triplecopyexplicit}, with again $F_{\mu\nu}$ replaced by $H_{\mu\nu}$ and the AdS metric used instead of the flat metric.
Note that these expressions, and their generalisation to higher spin, are  tensor forms of the spinor multicopy formula  given in
\cite{Didenko:2009td}.

The above results generalise to a Kerr black hole in AdS where it is useful to write the  AdS metric in spheroidal coordinates \cite{Didenko:2022qxq} 
\begin{equation}\label{eq:AdSKerrmetric}
  \bar g_{\mu\nu}dx^\mu dx^\nu = -W(1-\lambda r^2)dt^2 + F dr^2 + \frac{r^2+a^2\cos^2\!\theta}{1+\lambda a^2\cos^2\!\theta} d\theta^2 + \frac{(r^2+a^2) \sin^2\!\theta}{1+\lambda a^2} d\phi^2
    \, ,
\end{equation}
with $W=\frac{1+\lambda a^2\cos^2(\theta)}{1+\lambda a^2}$ and $F=\frac{r^2+a^2\cos^2(\theta)}{(1-\lambda r^2)(r^2+a^2)}$, and with the scalar field and Kerr-Schild vector now given by
\begin{equation}\label{eq:AdSKerrKerrSchild}
  \phi = \frac{2r}{r^2+a^2\cos^2\!\theta}\, , \quad k_\mu = \Big( F, W, 0,-\frac{a\sin^2\!\theta}{1+\lambda a^2} 
  \Big)
    \, .
\end{equation}
The equations of motion and field strengths are then given by the formul\ae\  presented above for AdS, but using the scalar and Kerr-Schild vector from \eqref{eq:AdSKerrKerrSchild}.


\subsection{Vector superspace}\label{sec: vector superspace}
Returning to $d$-dimensions, a vector superspace equation of motion in AdS space has been given in 
\cite{Segal:2001qq}, and generalised to fermionic fields in 
\cite{Najafizadeh:2018cpu} (along with correcting a factor in \cite{Segal:2001qq}). Both \cite{Segal:2001qq} are \cite{Najafizadeh:2018cpu} and based on the derivative 
\begin{equation}\label{eq:covderiv}
  \bar\nabla'_\mu = \partial_\mu + \Gamma^\rho_{\mu\nu}\eta_\rho\tilde\partial^\nu\, 
\end{equation}
with $\Gamma^\rho_{\mu\nu}$ the spacetime connection. This derivative fails to satisfy some expected identities such as $[\bar{\nabla}_\mu',g_{\rho\sigma}]A^{\alpha_1\ldots\alpha_n}=0$ and $[\bar{\nabla}_\mu',\eta_\nu]A^{\alpha_1\ldots\alpha_n}=0$, where $A^{\alpha_1\ldots\alpha_n}$ is any function of the spacetime coordinates and the $\eta_\mu$.

Let us instead define a covariant derivative as follows
\begin{equation}\label{eq: NewCovDer}
    \bar{\nabla}_\mu= \nabla_\mu+\Gamma^\rho_{\mu\nu}\eta_\rho\tilde\partial^\nu = \nabla_\mu-\nabla_\mu(\eta_\nu)\tilde{\partial}^\nu\, 
  \end{equation} 
(we assume that $\partial_\mu\eta_\nu=0$), where $\nabla_\mu$ is the usual spacetime covariant derivative for some metric $g_{\mu\nu}$. In flat space the $\eta_\mu$ vectors are simply coordinates in an additional vector space. However, in curved space it is natural to identify $\eta_\mu$ with coordinates on the cotangent bundle associated with a particular coordinate patch (as noted in \cite{Segal:2001qq}). To be concrete, given a chart with coordinates $x^\mu$ for some manifold $M$ there are canonical coordinates on the cotangent bundle $T^*M$ given by $(x,\eta)$ where
\begin{equation}
    (x,\eta)\rightarrow (x^\mu,\eta_\nu dx^\nu)\in T^* M\,.
\end{equation}
The two coordinates are independent so $\partial_\mu \eta_\nu=0$, and any particular fixed choice of $\eta_\mu$ defines a one-form (at least locally) to which we can apply the usual covariant derivative, as used in \eqref{eq: NewCovDer}. We will raise and lower indices on $\eta_\mu$ using the metric and, from the above definitions, under a coordinate transformation $x^\mu\rightarrow x^{\mu'}$ (and the induced transformation $\eta_\mu\rightarrow\eta_{\mu'}$) we have 
\begin{equation}
    \partial_{\mu'}= \frac{\partial x^{\mu}}{\partial x^{\mu'}}\partial_\mu+\left(\frac{\partial^2 x^{\nu'}}{\partial x^{\mu'}\partial x^\nu}\right)\eta_{\nu'}\tilde{\partial}_\nu\,,\quad
    \eta_{\mu'}=\frac{\partial x^\mu}{\partial x^{\mu'}} \eta_{\mu}, 
    \quad\tilde{\partial}^{\mu'}=\frac{\partial x^{\mu'}}{\partial x^{\mu}} \tilde{\partial}^{\mu} \implies \bar{\nabla}_{\mu'}=\frac{\partial x^\mu}{\partial x^{\mu'}}\bar{\nabla}_{\mu}\,.
\end{equation}
Note that the extra term in the transformation law of the spatial derivative $\partial_\mu$ implies that the usual derivative $\nabla_\mu$ does not transform covariantly when acting on a function of $x$ and $\eta$. Indeed, this is one motivation for introducing \eqref{eq: NewCovDer} in the first place (see Appendix \ref{app: cotangent} for more details).
It is also worth mentioning that \eqref{eq: NewCovDer} is useful even in flat space when using coordinates where $\Gamma^\rho_{\mu\nu}$ is non zero.
 As will be made clear in later examples, $\bar\nabla_\mu$ acts on expressions with indices in the usual way \textit{including} those indices on $\eta_\mu$ and $\tilde{\partial}^\mu$. (The definition  \eqref{eq: NewCovDer} may be compared with the natural metric on the co-tangent bundle, discussed in \cite{Gorbunov:2004na,10.2996/kmj/1138847443} for example.) 
This definition has several nice properties which we list below:\footnote{Equation \eqref{eq: torsion} indicates that there is torsion in this superspace, if we think of $\tilde\partial^\beta$ as the covariant derivative in the $\eta$ direction.}
\begin{align}
    &[\bar{\nabla}_\mu,g_{\rho\sigma}]A^{\alpha_1\ldots\alpha_n}=0\,,\\
    &[\bar{\nabla}_\mu,\eta_\nu]A^{\alpha_1\ldots\alpha_n}=0\,,\label{eq: comNewDeriv}\\
    &[\bar{\nabla}_\mu,\tilde{\partial}^\nu]A^{\alpha_1\ldots\alpha_n}=0\,,\\
    &[\bar{\nabla}_\mu,\bar{\nabla}_\nu]f=R^{\alpha}_{\,\beta\mu\nu}\eta_\alpha\tilde{\partial}^\beta f\,,\label{eq: torsion}\\
    &[\bar{\nabla}_\mu,\bar{\nabla}_\nu]X_\sigma=R^{\alpha}_{\,\beta\mu\nu}\eta_\alpha\tilde{\partial}^\beta X_\sigma+R_{\mu\nu\sigma}^{\quad\,\,\alpha}X_\alpha\,,\\
    &\bar{\nabla}_\mu (A^{\alpha_1\ldots\alpha_n} B^{\beta_1\ldots\beta_m})= (\bar{\nabla}_\mu A^{\alpha_1\ldots\alpha_n})B^{\beta_1\ldots\beta_m}+ A^{\alpha_1\ldots\alpha_n}(\bar{\nabla}_\mu B^{\beta_1\ldots\beta_m})\,,\\
    &\bar{\nabla}_\mu (\eta_{\alpha_1}\cdots\eta_{\alpha_n}A^{\alpha_1\ldots\alpha_n})=\eta_{\alpha_1}\cdots\eta_{\alpha_n}\bar{\nabla}_\mu A^{\alpha_1\ldots\alpha_n}\,,
\end{align}
where all of the tensors: $A^{\alpha_1\ldots}\,,f,X^\mu,Y^\mu$ are arbitrary functions of $x$ and $\eta$. If $\bar{\nabla}_\mu$ acts on an expression that does not depend on $\eta$ it simply reduces to the usual covariant derivative

Acting on a scalar superfield, one then has the identities
\begin{equation}
    \begin{split}  [\bar\Delta,\eta.\bar\nabla]&=\bar\nabla^2+R_{\,\,\beta}^{\alpha}\eta_\alpha\tilde{\partial}^\beta-R^{\alpha\,\,\mu}_{\,\,\beta\,\,\nu}\eta_\alpha\eta_\mu\tilde{\partial}^\beta\tilde{\partial}^\nu\,,\\ [\bar\nabla^2,\eta.\bar\nabla]&=R^{\mu\nu}\eta_\mu \bar{\nabla}_\nu+\nabla_\mu(R^{\alpha\,\,\mu\nu}_{\,\,\beta})\eta_\nu\eta_\alpha\tilde{\partial}^\beta+2R^{\alpha\,\,\mu\nu}_{\,\,\beta}\eta_\nu\eta_\alpha \tilde{\partial}^\beta\bar{\nabla}_\mu\,,\\
        [\bar\nabla^2,\bar\Delta]&=
        (\nabla^\nu {{R_{\nu\mu}}^\alpha}_\beta)\eta_\alpha\tilde\partial^\mu\tilde\partial^\beta - (\nabla^\nu R_{\nu\beta})\tilde\partial^\beta - R_{\mu\alpha}\bar\nabla^\alpha\tilde\partial^\mu + 2 {{R_{\nu\mu}}^\alpha}_\beta\eta_\alpha\bar\nabla^\nu\tilde\partial^\mu\tilde\partial^\beta\, ,
    \end{split}
\end{equation}
where $\bar\bigtriangleup = \tilde\partial. \bar\nabla$ and $\bar\nabla^2=\bar\nabla^\mu\bar\nabla_\mu$. In the AdS case these reduce to\footnote{In AdS space we have $R_{\mu\nu\rho\sigma}=\lambda(g_{\mu\rho}g_{\nu\sigma}-g_{\mu\sigma}g_{\nu\rho})$ and $R_{\mu\nu}=\lambda (d-1)g_{\mu\nu}$.}
\begin{equation}\begin{split}
[\bar\Delta,\eta.\bar\nabla]&=\bar\nabla^2+\lambda\big((d-1)\eta.\tilde\partial+ \eta^\mu\eta^\nu\tilde\partial_\mu\tilde\partial_\nu-\eta^2\tilde\partial^2\big), \\
[\bar\nabla^2,\eta.\bar\nabla]&=\lambda\big(2\eta.\bar\nabla\eta.\tilde\partial - 2 \eta^2\bar\Delta+(d-1)\eta.\bar\nabla\big), \\
[\bar\nabla^2,\bar\Delta]&=\lambda\big(2\eta.\bar\nabla\tilde\partial^2 - (d-1) \bar\Delta-2\eta.\tilde\partial\bar\Delta\big) \, .
\end{split}\end{equation}
Using these, one can show that the following equation of motion\footnote{This is equivalent to equation (2.25) in \cite{Najafizadeh:2018cpu} but is based on the covariant derivative \eqref{eq: NewCovDer} rather than \eqref{eq:covderiv}.}
\begin{equation}\label{eq:CSP_AdSeqmNEWTWO}
    \delta'(\eta^2+1)\bigg[\bar\nabla^2 \Psi -(\eta\cdot \bar\nabla)\bar\Delta\Psi +\frac{1}{2}(\eta^2+1)\bar\Delta^2 \Psi +
     \lambda \Big((\eta.\tilde\partial)^2 + (d-6)\eta.\tilde\partial - 2(d-3) + \eta^2\tilde\partial^2 + 2 \tilde\partial^2\Big)\Psi\bigg]=0\, ,
\end{equation}
is invariant under 
the  gauge transformations 
\begin{equation}\label{eq:AdSgauge_ivNEWTWO} 
 \delta_\epsilon\Psi  = \Big( \eta.\bar\nabla-\frac{1}{2}(\eta^2+1)\bar\Delta\Big)\epsilon +
  (\eta^2+1)^2\xi \, ,
\end{equation}
with the second-order gauge symmetry
\begin{equation}\label{eq:AdSgauge_ivNEW2TWO} 
 \delta\epsilon= (\eta^2+1)\Lambda,\quad \delta\xi = \frac{1}{2} \bar\Delta\Lambda \, .
\end{equation}
The equation of motion \eqref{eq:CSP_AdSeqmNEWTWO} can be obtained by varying the action \footnote{We thank Mojtaba Najafizadeh for asking if our equation of motion could be obtained from an action, which led us to add the explanation below.}
\begin{equation}\begin{split}\label{eq: action AdS}
    \int d^d x d^d \eta \Big[\,\frac{1}{2}\delta'(\eta^2+1)
    &
    (\bar{\nabla}_\mu\Psi) (\bar{\nabla}^\mu\Psi)+\frac{1}{4} \delta(\eta^2+1)(\bar{\Delta}\Psi)^2
    \\
    & +\delta'(\eta^2+1) \frac{\lambda}{2}\Psi \Big((\eta.\tilde\partial)^2 + (d-6)\eta.\tilde\partial - 2(d-3) + \eta^2\tilde\partial^2 + 2 \tilde\partial^2\Big)\Psi\Big]\,.
\end{split}\end{equation}
Note that since we are integrating over the cotangent bundle the integration measure no longer contains any factors of $\sqrt{|g|}$, as would be expected when just integrating over just the base manifold. This is due to the fact that, under a coordinate transformation $(x^\mu,\eta_\nu)\rightarrow(x^{\mu'},\eta_{\nu'})$
\begin{equation}
    dx^{\mu'}= \frac{\partial x^{\mu'}}{\partial x^{\mu}}dx^\mu\,, \quad 
    d\eta_{\mu'}=\frac{\partial x^\mu}{\partial x^{\mu'}}d \eta_\mu+\left(\frac{\partial^2 x^\nu}{\partial x^\mu \partial x^{\mu'}}\right)\eta_\nu dx^\mu\,.
\end{equation} 
When we wedge $dx$ and $d\eta$ the extra term in the transformation rule of $d\eta$ disappears and so the correct invariant measure is
\begin{equation}
    (\sqrt{|g|}d^dx)(\frac{1}{\sqrt{|g|}}d^d\eta)= d^d x d^d \eta\,,
\end{equation}
which is just the standard measure on phase space. This measure also has the property that the derivative \eqref{eq: NewCovDer} satisfies the usual integration-by-parts relations. To show this consider
\begin{equation}\label{eq: total der}
    \int d^d x d^d \eta\,\bar{\nabla}_\mu(X^\mu)= \int d^d x d^d \eta\,\nabla_\mu(X^\mu) - \int d^d x d^d \eta\,\nabla_\mu(\eta_\nu)\tilde{\partial}^\nu(X^\mu)\,,
\end{equation}
where $X^\mu$ is an arbitrary vector field depending on $x,\eta$.
The two terms above cancel each other, as we will now demonstrate. First we write
\begin{equation}
\begin{split}
    \int d^d x d^d \eta\,\nabla_\mu(\eta_\nu)\tilde{\partial}^\nu(X^\mu)&=\int d^d x d^d \eta\,\tilde{\partial}^\nu\left(\nabla_\mu(\eta_\nu)X^\mu \right) - 
    \int d^d x d^d \eta\,\tilde{\partial}^\nu\left(\nabla_\mu(\eta_\nu)\right)X^\mu\\
    &=  
    \int d^d x d^d \eta\,\Gamma^\nu_{\mu\nu}X^\mu\,,\\
    \end{split}
\end{equation}
where we have discarded a total derivative in $\eta$.
If we combine this with the first term in \eqref{eq: total der} we find 
\begin{equation}
    \int d^d x d^d \eta\,\bar{\nabla}_\mu(X^\mu)=  \int d^d x d^d \eta\, \nabla_\mu(X^\mu) -  \Gamma^\nu_{\mu\nu}X^\mu
    = \int d^d x d^d x \,\partial_\mu(X^\mu)=0
\end{equation}
where the result vanishes since it is a total spacetime derivative.
This gives the integration-by-parts identity
\begin{equation}
    \int d^d x d^d \eta\,\bar{\nabla}_\mu(X^\mu)=0 \implies  \int d^d x d^d \eta\,f\bar{\nabla}_\mu(X^\mu)=-\int d^d x d^d \eta\,\bar{\nabla}_\mu(f)X^\mu\,.
\end{equation}
With this one can show explicitly the equation of motion \eqref{eq:CSP_AdSeqmNEWTWO} can be obtained by varying the action \eqref{eq: action AdS}. Equivalently, one can also show that the operator in \eqref{eq:CSP_AdSeqmNEWTWO} is Hermitian.

We can now show that there is a generalised AdS Kerr-Schild solution analogous to the flat space expression \eqref{eq: expField} ie 
\begin{equation}\label{eq: expField2}
    \hat{\Psi}=f(k \cdot \eta)\phi\, ,
\end{equation}
where $k^2=0=k^\mu\nabla_\mu k^\nu$ here. Inserting this expression into \eqref{eq:CSP_AdSeqmNEWTWO}, and using similar manipulations to those below \eqref{eq: expField}, we find that these equations of motion are satisfied if
\begin{equation}\label{AdS012}
\begin{split}
&\nabla^2\phi -2(d-3)\lambda\phi = 0\, , \\
&\nabla^2(k^\mu\phi)-\nabla^\mu\nabla_\nu(k^\nu\phi) - (d-1)\lambda(k^\mu\phi) = 0\, , \\
&\nabla^2(k^\mu k^\nu\phi) - 2\nabla^{(\mu}\nabla_\rho(k^{\nu)}k^{\rho}\phi) - 2\lambda(k^\mu k^\nu\phi) = 0\, .  
\end{split}
\end{equation}
These are precisely the spin zero, one and two equations from \eqref{eq:AdS_HSeqnr} if one substitutes the Kerr-Schild fields $\phi\rightarrow\phi$, $\phi_\mu\rightarrow k_\mu\phi$ and $\phi_{\mu\nu}\rightarrow k_\mu k_\nu\phi$, with the $\phi$ on the right-hand sides the field in \eqref{eq: expField2}, satisfying the first equation in \eqref{AdS012}. If we choose $\hat{\Psi}= (k\cdot \eta)^s \phi$ we recover the result of \cite{Didenko:2022qxq} that the spin $s$ Kerr-Schild field satisfies the AdS higher spin equation of motion \eqref{eq:AdS_HSeqnr}, assuming the spin zero, one and two equations above are satisfied.

It is not clear if there is a continuous spin generalisation of \eqref{eq:CSP_AdSeqmNEWTWO} - if one replaces $\bar\Delta\rightarrow\bar\Delta+\rho$ in these equations of motion and in the gauge invariance \eqref{eq:AdSgauge_ivNEWTWO}, the order $\rho^3,\rho^2$ terms cancel in the variation of the equations of motion, but there is a non-cancelling order $\rho$ term
$ -\rho\lambda\delta'(\eta^2+1)(2\eta.\tilde\partial+d-1)(\eta^2+1)\epsilon$.
 We note that this contains no space-time derivatives, and so given the presence of these in the gauge transformations it appears unlikely that there is any local correction term that can be added to the equations of motion to obtain invariance.
 
Another way to explore this question is to consider the Pauli-Lubanski vector
\begin{equation}\label{eq:PLvector}
  W^\mu = \frac{1}{2}\epsilon^{\mu\nu\rho\sigma} J_{\mu\nu}P_\rho\, .
\end{equation}
In \cite{Schuster:2014hca} it was shown that for their flat space action and symmetries $W^2=-\rho^2$ up to gauge transformations, as required to describe a CSP (see also \cite{Rivelles:2016rwo} for a related discussion). This result does not appear to immediately generalise to AdS.

\section{Conclusions}\label{sec: conclusion}

In this paper we have discussed a generalisation of the Weyl double copy to higher spin "multi-copies", with the  linearised higher spin field strengths given by sums of powers of the Maxwell tensor.  This utilised  Kerr-Schild formulations of the fields. The Fronsdal equations for higher spin fields guarantee that the field strengths are traceless and divergence-free. Various examples were given. The spinor formulation provided further insights, working from the Penrose transform. The multi-copy was seen to provide information on admissible spacetime types under classification schemes. We studied the Schuster-Toro vector superspace formulation of the continuous spin particle, giving a generalised Kerr-Schild solution involving an arbitrary function. 
We then discussed the case of an anti-deSitter background 
and gave examples of the tensor multicopy, corresponding to the spinor formulations of \cite{Didenko:2009td}.
We clarified the vector superspace formulation of the theory, using a different definition of the covariant derivative,  with a generalised solution given in terms of the AdS-Kerr Kerr-Schild vector and scalar. We then described some of the obstacles to a continuous spin formulation.

Whether the multicopy generalises beyond linearised field strengths in flat space, or the AdS case, is one question. It would be interesting to study whether imposing some self-duality constraints does allow a multi-copy, notably using the spinor formalism and the Penrose transform on self-dual spaces, linking this for example to the discussion of the Eguchi-Hanson case in section \ref{sec: 4dPart1}.
Regarding nonlinear extensions, multicopy formul\ae\ involving corrections to the linearised higher spin black brane have been found in 
\cite{Didenko:2021vdb}.

A further question is the formulation of continuous spin particles (CSP) in non-flat spacetimes. The AdS formulation of the Schuster-Toro theory for $\rho=0$ does not give a CSP formulation upon shifting $\bar\bigtriangleup\rightarrow \bar\bigtriangleup+\rho$, as worked in flat space. However, continuous spin solutions in AdS have been studied in alternative approaches (see e.g. \cite{Metsaev:2021zdg,Metsaev:2025nbm,Buchbinder:2024vli}) and it would be interesting to see if these results can be translated  into the superspace formalism. It was interesting that the general function in \eqref{eq: expField} solves the higher spin equations and it may be possible to use this freedom to explore CSP solutions. However, in the case of CSPs the equations of motion for different spins will mix and thus the general function we found for $\rho=0$ may be constrained. It is also worth exploring possible CSP formulations on backgrounds such as the Schwarschild metric, although we found that solving the equations iteratively in $\rho$ did not prove possible. 


\vspace{24pt}

{\bf Acknowledgements:} 
We would like to thank Slava Didenko for helpful comments and for drawing our attention to some earlier work.
Thanks also to Ricardo Monteiro for  helpful remarks.
This work was supported by the Science and Technology Facilities Council (STFC) Consolidated Grants ST/P000754/1 “String theory, gauge theory and duality” and ST/T000686/1 “Amplitudes, strings and duality”. The work of G.R.B. is supported by the U.K. Royal Society through Grant URF{\textbackslash}R1 {\textbackslash}20109. No new data were generated or analysed during this work.

\appendix
\section{The cotangent bundle and covariant derivative}\label{app: cotangent}
Here we compile some useful properties about the cotangent bundle $T^*\!M$ and its symplectic and geometric structure which we used in the section \ref{sec: vector superspace} to define a covariant derivative. The material found here is covered in many standard textbooks e.g.\cite{SilvaSymplectic}. 

First let us restate our notation from the main text. Given a  manifold $M$, in each chart (with coordinates $x^\mu$) one can define natural coordinates on $T^*\!M$ given by ($x^\mu,\eta_\nu$)
\begin{equation}
    (x,\eta)\rightarrow (x^\mu,\eta_\nu dx^\nu)\in T^*\!M\,.
\end{equation}
From the definition of $\eta_\mu$ under a change of coordinates $x^\mu\rightarrow x^{\mu'}$ coordinates we have
\begin{equation}
    \eta_{\mu'}= \frac{\partial x^\mu}{\partial x^{\mu'}}\eta_\mu\,.
\end{equation}
The vectors $\partial_\mu=\frac{\partial}{\partial x^\mu}$ and $\tilde{\partial}^\mu=\frac{\partial}{\partial \eta_\mu}$ then naturally span the tangent space to $T^*\!M$. However, this basis of vectors does not transform in a nice way under a change of coordinates $(x^\mu,\eta^\mu)\rightarrow (x^{\mu'},\eta_{\mu'})$
\begin{equation}
    \partial_{\mu'}=\frac{\partial x^{\mu}}{\partial x^{\mu'}}\partial_\mu+\frac{\partial \eta_\mu}{\partial x^{\mu'}}\tilde{\partial}_\nu= \frac{\partial x^{\mu}}{\partial x^{\mu'}}\partial_\mu+\left(\frac{\partial^2 x^{\nu'}}{\partial x^{\mu'}\partial x^\nu}\right)\eta_{\nu'}\tilde{\partial}_\nu\,,\quad
    \quad\tilde{\partial}^{\mu'}=\frac{\partial x^{\mu'}}{\partial x^{\mu}} \tilde{\partial}^{\mu}\,.
\end{equation}
A better basis is given by a splitting into horizontal and vertical vectors
\begin{equation}
    H_\mu\coloneqq \partial_\mu + \Gamma_{\mu\nu}^\rho\eta_\rho \tilde{\partial}^\nu\,, \quad V^\mu\coloneqq \tilde{\partial}^\mu\,,
\end{equation}
which now have the simpler transformation properties
\begin{equation}
    H_{\mu'}= \frac{\partial x^{\mu}}{\partial x^{\mu'}} H_\mu\,,\quad V^{\mu'}= \frac{\partial x^{\mu'}}{\partial x^{\mu}} \tilde{\partial}^\mu\,.
\end{equation}
For the connection coefficients $\Gamma_{\mu\nu}^\rho$ we simply choose the usual Levi-Civita connection coefficients from $M$.
There is also a dual basis 
\begin{equation}
    \bar{H}^\mu\coloneqq dx^\mu\,,\quad \bar{V}_\mu\coloneqq d\eta_\mu-\Gamma_{\mu\nu}^\rho \eta_\rho dx^\nu  \,,
\end{equation}
which transforms similarly 
\begin{equation}
    \bar{V}_{\mu'}= \frac{\partial x^{\mu}}{\partial x^{\mu'}} \bar{V}_\mu\,,\quad \bar{H}^{\mu'}= \frac{\partial x^{\mu'}}{\partial x^{\mu}} \bar{H}^\mu\,,
\end{equation}
due to the transformation laws
\begin{equation}
    dx^{\mu'}= \frac{\partial x^{\mu'}}{\partial x^{\mu}}dx^\mu\,, \quad 
    d\eta_{\mu'}=\frac{\partial x^\mu}{\partial x^{\mu'}}d \eta_\mu+\left(\frac{\partial^2 x^\nu}{\partial x^\mu \partial x^{\mu'}}\right)\eta_\nu dx^\mu\,.
\end{equation} 

We extend the covariant derivative from $M$ to $T^*\!M$ so that it is compatible with horizontal and vertical splitting defined above 
\begin{equation}
    \nabla_{H_\mu}(H_\nu)\coloneqq \Gamma_{\mu\nu}^\rho H_\rho,\, \quad
    \nabla_{H_\mu}(\bar{V}_\mu)\coloneqq\Gamma_{\mu\nu}^\rho \bar{V}_\rho\,,\quad 
    \nabla_{H_\mu}(\bar{H}^\nu)\coloneqq -\Gamma_{\mu\rho}^\nu \bar{H}^\rho\,, \quad
    \nabla_{H_\mu}(V^\nu)\coloneqq -\Gamma_{\mu\rho}^\nu V^\rho\,,
\end{equation}
and 
\begin{equation}
     \nabla_{V^\mu}(H_\nu)=\nabla_{V^\mu}(\bar{V}_\mu)=0\,,\quad \nabla_{V^\mu}(\bar{H}^\nu)=\nabla_{V^\mu}(V^\nu)=0\,.
\end{equation}
With this choice $\nabla_{V^\mu}= \tilde{\partial}^\mu$ and $\nabla_{H_\mu}= \bar{\nabla}_\mu$ defined in the main text. For example
\begin{equation}
\nabla_{H_\mu} (X^\nu H_\nu)= (H_{\mu}(X^\nu) + \Gamma_{\mu\rho}^{\nu}X^\rho)H_\nu = (\partial_{\mu}(X^\nu) +\Gamma_{\mu \sigma}^\rho \eta_\rho \tilde{\partial}^\sigma X^\mu+ \Gamma_{\mu\rho}^{\nu}X^\rho)H_\nu\,.    
\end{equation}

This covariant derivative also preserves the  canonical symplectic form
\begin{equation}
    \omega \coloneqq d\eta_\mu \wedge d x^\mu= \bar{V}_\mu \wedge \bar{H}^\mu\,,\quad \nabla_{H_\mu}\omega=\nabla_{V^\mu}\omega=0
\end{equation}
which looks the same in any choice of coordinates, and in the second equality above we have used the torsionless condition of $\Gamma_{\mu\nu}^\rho$. We can use this to relate the horizontal/vertical basis and its dual 
\begin{equation}
    \omega(V^\mu,\_)= \bar{H}^\mu\,,\quad \omega(\_,H^\mu)=\bar{V}^\mu\,.
\end{equation}

The metric on the manifold $M$ can also be lifted to a Sasaki metric \cite{10.2748/tmj/1178244668}\footnote{The Sasaki metric was originally defined for $TM$ not $T^*\!M$ but the structure is essentially the same.} on $T^*\!M$ (see also \cite{Gorbunov:2004na,10.2996/kmj/1138847443}) given by
\begin{equation}
    ds^2_{T^*\!M}= g_{\mu\nu} (x)\bar{H}^\mu\bar{H}^\nu+ g^{\mu\nu}(x)\bar{V}_\mu \bar{V}_\nu\,.
\end{equation}
With this definition the metric components $g_{\mu\nu}$ and its inverse $g^{\mu\nu}$ transform in the usual way. In addition, the metric above has vanishing covariant derivative as expected since the connection coefficients $\Gamma_{\mu\nu}^\rho$ come from the Levi-Civita connection on $M$. Due to the compatibility of the covariant derivative with the symplectic form and the metric, we will write component fields as $X^\mu$ without specifying whether they belong to a vector or covector. Indeed, $\nabla X^\mu$ takes the same value regardless of whether $X^\mu$ was the component of $X^\mu H_\mu$ or $X^\mu \bar{V}_\mu$, and similarly for $X_\mu$. For concreteness, one can exclusively use either the horizontal basis or the vertical one.


\bibliographystyle{utphys}
\bibliography{remainder}

\providecommand{\href}[2]{#2}\begingroup\raggedright\begin{thebibliography}{10}

\bibitem{Bern:2008qj}
Z.~Bern, J.~J.~M. Carrasco, and H.~Johansson, ``{New Relations for Gauge-Theory Amplitudes},'' \href{http://dx.doi.org/10.1103/PhysRevD.78.085011}{{\em Phys. Rev. D} {\bfseries 78} (2008) 085011}, \href{http://arxiv.org/abs/0805.3993}{{\ttfamily arXiv:0805.3993 [hep-ph]}}.

\bibitem{Bern:2010ue}
Z.~Bern, J.~J.~M. Carrasco, and H.~Johansson, ``{Perturbative Quantum Gravity as a Double Copy of Gauge Theory},'' \href{http://dx.doi.org/10.1103/PhysRevLett.105.061602}{{\em Phys. Rev. Lett.} {\bfseries 105} (2010) 061602}, \href{http://arxiv.org/abs/1004.0476}{{\ttfamily arXiv:1004.0476 [hep-th]}}.

\bibitem{Travaglini:2022uwo}
G.~Travaglini {\em et~al.}, ``{The SAGEX review on scattering amplitudes},'' \href{http://dx.doi.org/10.1088/1751-8121/ac8380}{{\em J. Phys. A} {\bfseries 55} no.~44, (2022) 443001}, \href{http://arxiv.org/abs/2203.13011}{{\ttfamily arXiv:2203.13011 [hep-th]}}.

\bibitem{White:2024pve}
C.~D. White, \href{http://dx.doi.org/10.1142/q0457}{{\em {The Classical Double Copy}}}.
\newblock World Scientific, 5, 2024.

\bibitem{Luna:2018dpt}
A.~Luna, R.~Monteiro, I.~Nicholson, and D.~O'Connell, ``{Type D Spacetimes and the Weyl Double Copy},'' \href{http://dx.doi.org/10.1088/1361-6382/ab03e6}{{\em Class. Quant. Grav.} {\bfseries 36} (2019) 065003}, \href{http://arxiv.org/abs/1810.08183}{{\ttfamily arXiv:1810.08183 [hep-th]}}.

\bibitem{WalkerPenrose}
M.~Walker and R.~Penrose, ``{On quadratic first integrals of the geodesic equations for type [22] spacetimes},'' \href{http://dx.doi.org/10.1007/BF01649445}{{\em Commun. Math. Phys.} {\bfseries 18} (1970) 265--274}.

\bibitem{Hughston}
L.~Hughston, R.~Penrose, P.~Sommers, and M.~Walker, ``{On a quadratic first integral for the charged particle orbits in the charged Kerr solution},'' \href{http://dx.doi.org/10.1007/BF01645517}{{\em Commun. Math. Phys.} {\bfseries 27} (1972) 303--30}.

\bibitem{Dietz}
W.~Dietz and R.~R{\"u}diger, ``{Space-times admitting Killing-Yano tensors},'' \href{http://dx.doi.org/10.1098/rspa.1981. 0056}{{\em Proc. R. Soc. Lond. A} {\bfseries 375} (1981) 361--378}.

\bibitem{Godazgar:2020zbv}
H.~Godazgar, M.~Godazgar, R.~Monteiro, D.~Peinador~Veiga, and C.~N. Pope, ``{Weyl Double Copy for Gravitational Waves},'' \href{http://dx.doi.org/10.1103/PhysRevLett.126.101103}{{\em Phys. Rev. Lett.} {\bfseries 126} no.~10, (2021) 101103}, \href{http://arxiv.org/abs/2010.02925}{{\ttfamily arXiv:2010.02925 [hep-th]}}.

\bibitem{Sabharwal:2019ngs}
S.~Sabharwal and J.~W. Dalhuisen, ``{Anti-Self-Dual Spacetimes, Gravitational Instantons and Knotted Zeros of the Weyl Tensor},'' \href{http://dx.doi.org/10.1007/JHEP07(2019)004}{{\em JHEP} {\bfseries 07} (2019) 004}, \href{http://arxiv.org/abs/1904.06030}{{\ttfamily arXiv:1904.06030 [hep-th]}}.

\bibitem{Alawadhi:2019urr}
R.~Alawadhi, D.~S. Berman, B.~Spence, and D.~Peinador~Veiga, ``{S-duality and the double copy},'' \href{http://dx.doi.org/10.1007/JHEP03(2020)059}{{\em JHEP} {\bfseries 03} (2020) 059}, \href{http://arxiv.org/abs/1911.06797}{{\ttfamily arXiv:1911.06797 [hep-th]}}.

\bibitem{Alawadhi:2020jrv}
R.~Alawadhi, D.~S. Berman, and B.~Spence, ``{Weyl doubling},'' \href{http://dx.doi.org/10.1007/JHEP09(2020)127}{{\em JHEP} {\bfseries 09} (2020) 127}, \href{http://arxiv.org/abs/2007.03264}{{\ttfamily arXiv:2007.03264 [hep-th]}}.

\bibitem{Elor:2020nqe}
G.~Elor, K.~Farnsworth, M.~L. Graesser, and G.~Herczeg, ``{The Newman-Penrose Map and the Classical Double Copy},'' \href{http://dx.doi.org/10.1007/JHEP12(2020)121}{{\em JHEP} {\bfseries 12} (2020) 121}, \href{http://arxiv.org/abs/2006.08630}{{\ttfamily arXiv:2006.08630 [hep-th]}}.

\bibitem{Chacon:2021wbr}
E.~Chac\'on, S.~Nagy, and C.~D. White, ``{The Weyl double copy from twistor space},'' \href{http://dx.doi.org/10.1007/JHEP05(2021)239}{{\em JHEP} {\bfseries 05} (2021) 2239}, \href{http://arxiv.org/abs/2103.16441}{{\ttfamily arXiv:2103.16441 [hep-th]}}.

\bibitem{Chacon:2021lox}
E.~Chac\'on, S.~Nagy, and C.~D. White, ``{Alternative formulations of the twistor double copy},'' \href{http://dx.doi.org/10.1007/JHEP03(2022)180}{{\em JHEP} {\bfseries 03} (2022) 180}, \href{http://arxiv.org/abs/2112.06764}{{\ttfamily arXiv:2112.06764 [hep-th]}}.

\bibitem{Schuster:2014hca}
P.~Schuster and N.~Toro, ``{Continuous-spin particle field theory with helicity correspondence},'' \href{http://dx.doi.org/10.1103/PhysRevD.91.025023}{{\em Phys. Rev. D} {\bfseries 91} (2015) 025023}, \href{http://arxiv.org/abs/1404.0675}{{\ttfamily arXiv:1404.0675 [hep-th]}}.

\bibitem{KerrSchild1}
R.~Kerr and A.~Schild, ``{A new class of vacuum solutions of the Einstein field equations, in G. Barbera (ed.), "Atti del convegno sulla relativita` generale; problemi dell’energia e onde gravitazionali"},''  (1965) 173.

\bibitem{KerrSchild2}
R.~Kerr and A.~Schild, ``{Some algebraically degenerate solutions of Einstein’s gravitational field equations},'' {\em R. Finn (ed.), Proc. Sym. in Applied Math} {\bfseries XVII} (1965) 199–209.

\bibitem{MacCallum}
H.~Stephani, D.~Kramer, M.~MacCallum, C.~Hoenselaers, and E.~Herlt, \href{http://dx.doi.org/10.1017/CBO9780511535185}{{\em {Exact solutions of Einstein’s field equations}}}.
\newblock Cambridge University Press, 2003.

\bibitem{Monteiro:2014cda}
R.~Monteiro, D.~O'Connell, and C.~D. White, ``{Black holes and the double copy},'' \href{http://dx.doi.org/10.1007/JHEP12(2014)056}{{\em JHEP} {\bfseries 12} (2014) 056}, \href{http://arxiv.org/abs/1410.0239}{{\ttfamily arXiv:1410.0239 [hep-th]}}.

\bibitem{Luna:2015paa}
A.~Luna, R.~Monteiro, D.~O'Connell, and C.~D. White, ``{The classical double copy for Taub\textendash{}NUT spacetime},'' \href{http://dx.doi.org/10.1016/j.physletb.2015.09.021}{{\em Phys. Lett. B} {\bfseries 750} (2015) 272--277}, \href{http://arxiv.org/abs/1507.01869}{{\ttfamily arXiv:1507.01869 [hep-th]}}.

\bibitem{Didenko:2008va}
V.~E. Didenko, A.~S. Matveev, and M.~A. Vasiliev, ``{Unfolded Description of AdS(4) Kerr Black Hole},'' \href{http://dx.doi.org/10.1016/j.physletb.2008.05.067}{{\em Phys. Lett. B} {\bfseries 665} (2008) 284--293}, \href{http://arxiv.org/abs/0801.2213}{{\ttfamily arXiv:0801.2213 [gr-qc]}}.

\bibitem{Didenko:2009td}
V.~E. Didenko and M.~A. Vasiliev, ``{Static BPS black hole in 4d higher-spin gauge theory},'' \href{http://dx.doi.org/10.1016/j.physletb.2009.11.023}{{\em Phys. Lett. B} {\bfseries 682} (2009) 305--315}, \href{http://arxiv.org/abs/0906.3898}{{\ttfamily arXiv:0906.3898 [hep-th]}}. [Erratum: Phys.Lett.B 722, 389 (2013)].

\bibitem{Didenko:2022qxq}
V.~E. Didenko and N.~K. Dosmanbetov, ``{Classical Double Copy and Higher-Spin Fields},'' \href{http://dx.doi.org/10.1103/PhysRevLett.130.071603}{{\em Phys. Rev. Lett.} {\bfseries 130} no.~7, (2023) 071603}, \href{http://arxiv.org/abs/2210.04704}{{\ttfamily arXiv:2210.04704 [hep-th]}}.

\bibitem{Ponomarev:2022vjb}
D.~Ponomarev, ``{Basic Introduction to Higher-Spin Theories},'' \href{http://dx.doi.org/10.1007/s10773-023-05399-5}{{\em Int. J. Theor. Phys.} {\bfseries 62} no.~7, (2023) 146}, \href{http://arxiv.org/abs/2206.15385}{{\ttfamily arXiv:2206.15385 [hep-th]}}.

\bibitem{Sorokin:2004ie}
D.~Sorokin, ``{Introduction to the classical theory of higher spins},'' \href{http://dx.doi.org/10.1063/1.1923335}{{\em AIP Conf. Proc.} {\bfseries 767} no.~1, (2005) 172--202}, \href{http://arxiv.org/abs/hep-th/0405069}{{\ttfamily arXiv:hep-th/0405069}}.

\bibitem{Bouatta:2004kk}
N.~Bouatta, G.~Compere, and A.~Sagnotti, ``{An Introduction to free higher-spin fields},'' in {\em {1st Solvay Workshop on Higher Spin Gauge Theories}}, pp.~79--99.
\newblock 9, 2004.
\newblock \href{http://arxiv.org/abs/hep-th/0409068}{{\ttfamily arXiv:hep-th/0409068}}.

\bibitem{deMedeiros:2002qpr}
P.~de~Medeiros and C.~Hull, ``{Exotic tensor gauge theory and duality},'' \href{http://dx.doi.org/10.1007/s00220-003-0810-z}{{\em Commun. Math. Phys.} {\bfseries 235} (2003) 255--273}, \href{http://arxiv.org/abs/hep-th/0208155}{{\ttfamily arXiv:hep-th/0208155}}.

\bibitem{deMedeiros:2003osq}
P.~de~Medeiros and C.~Hull, ``{Geometric second order field equations for general tensor gauge fields},'' \href{http://dx.doi.org/10.1088/1126-6708/2003/05/019}{{\em JHEP} {\bfseries 05} (2003) 019}, \href{http://arxiv.org/abs/hep-th/0303036}{{\ttfamily arXiv:hep-th/0303036}}.

\bibitem{White:2020sfn}
C.~D. White, ``{Twistorial Foundation for the Classical Double Copy},'' \href{http://dx.doi.org/10.1103/PhysRevLett.126.061602}{{\em Phys. Rev. Lett.} {\bfseries 126} no.~6, (2021) 061602}, \href{http://arxiv.org/abs/2012.02479}{{\ttfamily arXiv:2012.02479 [hep-th]}}.

\bibitem{Armstrong-Williams:2024bog}
K.~Armstrong-Williams, N.~Moynihan, and C.~D. White, ``{Deriving Weyl double copies with sources},'' \href{http://arxiv.org/abs/2407.18107}{{\ttfamily arXiv:2407.18107 [hep-th]}}.

\bibitem{Didenko:2021vui}
V.~E. Didenko and A.~V. Korybut, ``{Planar solutions of higher-spin theory. Part I. Free field level},'' \href{http://dx.doi.org/10.1007/JHEP08(2021)144}{{\em JHEP} {\bfseries 08} (2021) 144}, \href{http://arxiv.org/abs/2105.09021}{{\ttfamily arXiv:2105.09021 [hep-th]}}.

\bibitem{Monteiro:2018xev}
R.~Monteiro, I.~Nicholson, and D.~O'Connell, ``{Spinor-helicity and the algebraic classification of higher-dimensional spacetimes},'' \href{http://dx.doi.org/10.1088/1361-6382/ab03df}{{\em Class. Quant. Grav.} {\bfseries 36} (2019) 065006}, \href{http://arxiv.org/abs/1809.03906}{{\ttfamily arXiv:1809.03906 [gr-qc]}}.

\bibitem{Wigner}
E.~Wigner, ``{On Unitary Representations of the Inhomogeneous Lorentz Group},'' {\em Annals of Mathematics} {\bfseries 40} (1939) 149.

\bibitem{Rivelles:2014fsa}
V.~O. Rivelles, ``{Gauge Theory Formulations for Continuous and Higher Spin Fields},'' \href{http://dx.doi.org/10.1103/PhysRevD.91.125035}{{\em Phys. Rev. D} {\bfseries 91} no.~12, (2015) 125035}, \href{http://arxiv.org/abs/1408.3576}{{\ttfamily arXiv:1408.3576 [hep-th]}}.

\bibitem{Rivelles:2016rwo}
V.~O. Rivelles, ``{Remarks on a Gauge Theory for Continuous Spin Particles},'' \href{http://dx.doi.org/10.1140/epjc/s10052-017-4927-1}{{\em Eur. Phys. J. C} {\bfseries 77} no.~7, (2017) 433}, \href{http://arxiv.org/abs/1607.01316}{{\ttfamily arXiv:1607.01316 [hep-th]}}.

\bibitem{Bekaert:2017khg}
X.~Bekaert and E.~D. Skvortsov, ``{Elementary particles with continuous spin},'' \href{http://dx.doi.org/10.1142/S0217751X17300198}{{\em Int. J. Mod. Phys. A} {\bfseries 32} no.~23n24, (2017) 1730019}, \href{http://arxiv.org/abs/1708.01030}{{\ttfamily arXiv:1708.01030 [hep-th]}}.

\bibitem{Najafizadeh:2017tin}
M.~Najafizadeh, ``{Modified Wigner equations and continuous spin gauge field},'' \href{http://dx.doi.org/10.1103/PhysRevD.97.065009}{{\em Phys. Rev. D} {\bfseries 97} no.~6, (2018) 065009}, \href{http://arxiv.org/abs/1708.00827}{{\ttfamily arXiv:1708.00827 [hep-th]}}.

\bibitem{Schuster:2014xja}
P.~Schuster and N.~Toro, ``{A new class of particle in 2 + 1 dimensions},'' \href{http://dx.doi.org/10.1016/j.physletb.2015.02.050}{{\em Phys. Lett. B} {\bfseries 743} (2015) 224--227}, \href{http://arxiv.org/abs/1404.1076}{{\ttfamily arXiv:1404.1076 [hep-th]}}.

\bibitem{Bekaert:2015qkt}
X.~Bekaert, M.~Najafizadeh, and M.~R. Setare, ``{A gauge field theory of fermionic Continuous-Spin Particles},'' \href{http://dx.doi.org/10.1016/j.physletb.2016.07.005}{{\em Phys. Lett. B} {\bfseries 760} (2016) 320--323}, \href{http://arxiv.org/abs/1506.00973}{{\ttfamily arXiv:1506.00973 [hep-th]}}.

\bibitem{Najafizadeh:2019mun}
M.~Najafizadeh, ``{Supersymmetric Continuous Spin Gauge Theory},'' \href{http://dx.doi.org/10.1007/JHEP03(2020)027}{{\em JHEP} {\bfseries 03} (2020) 027}, \href{http://arxiv.org/abs/1912.12310}{{\ttfamily arXiv:1912.12310 [hep-th]}}.

\bibitem{Najafizadeh:2021dsm}
M.~Najafizadeh, ``{Off-shell supersymmetric continuous spin gauge theory},'' \href{http://dx.doi.org/10.1007/JHEP02(2022)038}{{\em JHEP} {\bfseries 02} (2022) 038}, \href{http://arxiv.org/abs/2112.10178}{{\ttfamily arXiv:2112.10178 [hep-th]}}.

\bibitem{Schuster:2023jgc}
P.~Schuster and N.~Toro, ``{Quantum electrodynamics mediated by a photon with continuous spin},'' \href{http://dx.doi.org/10.1103/PhysRevD.109.096008}{{\em Phys. Rev. D} {\bfseries 109} no.~9, (2024) 096008}, \href{http://arxiv.org/abs/2308.16218}{{\ttfamily arXiv:2308.16218 [hep-th]}}.

\bibitem{Metsaev:2018lth}
R.~R. Metsaev, ``{BRST-BV approach to continuous-spin field},'' \href{http://dx.doi.org/10.1016/j.physletb.2018.04.038}{{\em Phys. Lett. B} {\bfseries 781} (2018) 568--573}, \href{http://arxiv.org/abs/1803.08421}{{\ttfamily arXiv:1803.08421 [hep-th]}}.

\bibitem{Buchbinder:2018yoo}
I.~L. Buchbinder, V.~A. Krykhtin, and H.~Takata, ``{BRST approach to Lagrangian construction for bosonic continuous spin field},'' \href{http://dx.doi.org/10.1016/j.physletb.2018.07.070}{{\em Phys. Lett. B} {\bfseries 785} (2018) 315--319}, \href{http://arxiv.org/abs/1806.01640}{{\ttfamily arXiv:1806.01640 [hep-th]}}.

\bibitem{Alkalaev:2017hvj}
K.~B. Alkalaev and M.~A. Grigoriev, ``{Continuous spin fields of mixed-symmetry type},'' \href{http://dx.doi.org/10.1007/JHEP03(2018)030}{{\em JHEP} {\bfseries 03} (2018) 030}, \href{http://arxiv.org/abs/1712.02317}{{\ttfamily arXiv:1712.02317 [hep-th]}}.

\bibitem{Schuster:2023xqa}
P.~Schuster, N.~Toro, and K.~Zhou, ``{Interactions of Particles with ''Continuous Spin'' Fields},'' \href{http://dx.doi.org/10.1007/JHEP04(2023)010}{{\em JHEP} {\bfseries 04} (2023) 010}, \href{http://arxiv.org/abs/2303.04816}{{\ttfamily arXiv:2303.04816 [hep-th]}}.

\bibitem{Bellazzini:2024dco}
B.~Bellazzini, S.~De~Angelis, and M.~Romano, ``{Continuous-Spin Particles, On Shell},'' \href{http://arxiv.org/abs/2406.17017}{{\ttfamily arXiv:2406.17017 [hep-th]}}.

\bibitem{Schuster:2024wjc}
P.~Schuster, G.~Sundaresan, and N.~Toro, ``{Thermodynamics of continuous spin photons},'' \href{http://dx.doi.org/10.1103/PhysRevD.111.056019}{{\em Phys. Rev. D} {\bfseries 111} no.~5, (2025) 056019}, \href{http://arxiv.org/abs/2406.14616}{{\ttfamily arXiv:2406.14616 [hep-ph]}}.

\bibitem{Reilly:2025lnm}
A.~Reilly, P.~Schuster, and N.~Toro, ``{Probing ''Continuous Spin'' QED with Rare Atomic Transitions},'' \href{http://arxiv.org/abs/2505.01500}{{\ttfamily arXiv:2505.01500 [hep-ph]}}.

\bibitem{Kundu:2025fsd}
S.~Kundu, P.~Schuster, and N.~Toro, ``{A First Look at ''Continuous Spin'' Gravity -- Time Delay Signatures},'' \href{http://arxiv.org/abs/2503.03817}{{\ttfamily arXiv:2503.03817 [gr-qc]}}.

\bibitem{Metsaev:2025qkr}
R.~R. Metsaev, ``{Interacting massive/massless continuous-spin fields and integer-spin fields},'' \href{http://arxiv.org/abs/2505.02817}{{\ttfamily arXiv:2505.02817 [hep-th]}}.

\bibitem{Kundu:2025mzm}
S.~Kundu, A.~Russo, P.~Schuster, and N.~Toro, ``{Interactions of a Continuous-Spin Field with a Spin-1/2 Particle},'' \href{http://arxiv.org/abs/2505.14770}{{\ttfamily arXiv:2505.14770 [hep-th]}}.

\bibitem{Reilly:2025vqs}
A.~Reilly, A.~Russo, P.~Schuster, and N.~Toro, ``{Hydrogen 21 cm Constraints on the Photon's Spin Scale},'' \href{http://arxiv.org/abs/2505.15890}{{\ttfamily arXiv:2505.15890 [hep-ph]}}.

\bibitem{Metsaev:2016lhs}
R.~R. Metsaev, ``{Continuous spin gauge field in (A)dS space},'' \href{http://dx.doi.org/10.1016/j.physletb.2017.02.027}{{\em Phys. Lett. B} {\bfseries 767} (2017) 458--464}, \href{http://arxiv.org/abs/1610.00657}{{\ttfamily arXiv:1610.00657 [hep-th]}}.

\bibitem{Metsaev:2017ytk}
R.~R. Metsaev, ``{Fermionic continuous spin gauge field in (A)dS space},'' \href{http://dx.doi.org/10.1016/j.physletb.2017.08.020}{{\em Phys. Lett. B} {\bfseries 773} (2017) 135--141}, \href{http://arxiv.org/abs/1703.05780}{{\ttfamily arXiv:1703.05780 [hep-th]}}.

\bibitem{Metsaev:2017myp}
R.~R. Metsaev, ``{Continuous-spin mixed-symmetry fields in AdS(5)},'' \href{http://dx.doi.org/10.1088/1751-8121/aabcda}{{\em J. Phys. A} {\bfseries 51} no.~21, (2018) 215401}, \href{http://arxiv.org/abs/1711.11007}{{\ttfamily arXiv:1711.11007 [hep-th]}}.

\bibitem{Khabarov:2017lth}
M.~V. Khabarov and Y.~M. Zinoviev, ``{Infinite (continuous) spin fields in the frame-like formalism},'' \href{http://dx.doi.org/10.1016/j.nuclphysb.2018.01.016}{{\em Nucl. Phys. B} {\bfseries 928} (2018) 182--216}, \href{http://arxiv.org/abs/1711.08223}{{\ttfamily arXiv:1711.08223 [hep-th]}}.

\bibitem{Metsaev:2019opn}
R.~R. Metsaev, ``{Light-cone continuous-spin field in AdS space},'' \href{http://dx.doi.org/10.1016/j.physletb.2019.04.041}{{\em Phys. Lett. B} {\bfseries 793} (2019) 134--140}, \href{http://arxiv.org/abs/1903.10495}{{\ttfamily arXiv:1903.10495 [hep-th]}}.

\bibitem{Metsaev:2021zdg}
R.~R. Metsaev, ``{Mixed-symmetry continuous-spin fields in flat and AdS spaces},'' \href{http://dx.doi.org/10.1016/j.physletb.2021.136497}{{\em Phys. Lett. B} {\bfseries 820} (2021) 136497}, \href{http://arxiv.org/abs/2105.11281}{{\ttfamily arXiv:2105.11281 [hep-th]}}.

\bibitem{Buchbinder:2024hea}
I.~L. Buchbinder, S.~A. Fedoruk, A.~P. Isaev, and V.~A. Krykhtin, ``{Infinite (continuous) spin particle in constant curvature space},'' \href{http://dx.doi.org/10.1016/j.physletb.2024.138689}{{\em Phys. Lett. B} {\bfseries 853} (2024) 138689}, \href{http://arxiv.org/abs/2402.13879}{{\ttfamily arXiv:2402.13879 [hep-th]}}.

\bibitem{Buchbinder:2024jpt}
I.~L. Buchbinder, S.~A. Fedoruk, A.~P. Isaev, and V.~A. Krykhtin, ``{BRST construction for infinite spin field on $AdS_4$},'' \href{http://dx.doi.org/10.1140/epjp/s13360-024-05430-6}{{\em Eur. Phys. J. Plus} {\bfseries 139} no.~7, (2024) 621}, \href{http://arxiv.org/abs/2403.14446}{{\ttfamily arXiv:2403.14446 [hep-th]}}.

\bibitem{Didenko:2011ir}
V.~E. Didenko, ``{Coordinate independent approach to 5d black holes},'' \href{http://dx.doi.org/10.1088/0264-9381/29/2/025009}{{\em Class. Quant. Grav.} {\bfseries 29} (2012) 025009}, \href{http://arxiv.org/abs/1108.4321}{{\ttfamily arXiv:1108.4321 [hep-th]}}.

\bibitem{Segal:2001qq}
A.~Y. Segal, ``{A Generating formulation for free higher spin massless fields},'' \href{http://arxiv.org/abs/hep-th/0103028}{{\ttfamily arXiv:hep-th/0103028}}.

\bibitem{Najafizadeh:2018cpu}
M.~Najafizadeh, ``{Local action for fermionic unconstrained higher spin gauge fields in AdS and dS spacetimes},'' \href{http://dx.doi.org/10.1103/PhysRevD.98.125012}{{\em Phys. Rev. D} {\bfseries 98} no.~12, (2018) 125012}, \href{http://arxiv.org/abs/1807.01124}{{\ttfamily arXiv:1807.01124 [hep-th]}}.

\bibitem{Gorbunov:2004na}
I.~V. Gorbunov, S.~L. Lyakhovich, and A.~A. Sharapov, ``{Wick quantization of cotangent bundles over Riemannian manifolds},'' \href{http://dx.doi.org/10.1016/j.geomphys.2004.06.003}{{\em J. Geom. Phys.} {\bfseries 53} (2005) 98--121}, \href{http://arxiv.org/abs/hep-th/0401022}{{\ttfamily arXiv:hep-th/0401022}}.

\bibitem{10.2996/kmj/1138847443}
K.~P. Mok, ``{Metrics and connections on the cotangent bundle},'' \href{http://dx.doi.org/10.2996/kmj/1138847443}{{\em Kodai Mathematical Seminar Reports} {\bfseries 28} no.~2-3, (1977) 226 -- 238}. \url{https://doi.org/10.2996/kmj/1138847443}.

\bibitem{Didenko:2021vdb}
V.~E. Didenko and A.~V. Korybut, ``{Planar solutions of higher-spin theory. Nonlinear corrections},'' \href{http://dx.doi.org/10.1007/JHEP01(2022)125}{{\em JHEP} {\bfseries 01} (2022) 125}, \href{http://arxiv.org/abs/2110.02256}{{\ttfamily arXiv:2110.02256 [hep-th]}}.

\bibitem{Metsaev:2025nbm}
R.~R. Metsaev, ``{Light-cone vector superspace and continuous-spin field in AdS},'' \href{http://arxiv.org/abs/2507.05194}{{\ttfamily arXiv:2507.05194 [hep-th]}}.

\bibitem{Buchbinder:2024vli}
I.~L. Buchbinder, S.~A. Fedoruk, A.~P. Isaev, and M.~A. Podoinitsyn, ``{On the realization of infinite (continuous) spin field representations in AdS4 space},'' \href{http://dx.doi.org/10.1016/j.physletb.2024.139226}{{\em Phys. Lett. B} {\bfseries 861} (2025) 139226}, \href{http://arxiv.org/abs/2410.07873}{{\ttfamily arXiv:2410.07873 [hep-th]}}.

\bibitem{SilvaSymplectic}
A.~Cannas~da Silva, \href{http://dx.doi.org/10.1007/978-3-540-45330-7}{{\em { Lectures on Symplectic Geometry}}}.
\newblock Springer Berlin, Heidelberg, 2001.

\bibitem{10.2748/tmj/1178244668}
S.~Sasaki, ``{On the differential geometry of tangent bundles of Riemannian manifolds},'' \href{http://dx.doi.org/10.2748/tmj/1178244668}{{\em Tohoku Mathematical Journal} {\bfseries 10} no.~3, (1958) 338 -- 354}. \url{https://doi.org/10.2748/tmj/1178244668}.

\end{thebibliography}\endgroup
\end{document}